\documentclass[]{article}

\usepackage[utf8]{inputenc}
\usepackage{booktabs}
\usepackage{graphicx}
\usepackage{array}
\usepackage[showframe=false]{geometry}
\usepackage{changepage}
\usepackage[]{graphicx}
\usepackage{caption}
\usepackage{subcaption}
\usepackage[toc,page]{appendix}
\usepackage{bm}
\usepackage{float}
\usepackage{amsmath}
\usepackage{mathtools}
\usepackage{authblk}
\usepackage[colorlinks=true, allcolors=blue]{hyperref}
\usepackage{longtable}
\numberwithin{equation}{section}
\usepackage[dvipsnames]{xcolor}
\usepackage{afterpage}
\usepackage{makecell}
\usepackage{amssymb}
\usepackage{hyperref}
\usepackage{cite}
\usepackage{cleveref}
\usepackage{lmodern}
\usepackage{hyperref}

\crefname{equation}{Eq.}{Eqs.}
\crefrangelabelformat{equation}{(#3#1#4)--(#5#2#6)}

\newcommand{\res}[1]{Section~\ref{#1}}

\newcommand{\refig}[1]{Fig.~\ref{#1}}
\newcommand{\ret}[1]{Table~\ref{#1}}

\begin{document}





  \vspace*{-30truemm}
  \begin{flushright}
    MITP-25-079\\
  \end{flushright}
  \vspace{15truemm}


  \vskip 10 true mm
  \centerline{\Large \textbf{$f_K/f_{\pi}$ in iso-symmetric QCD and the CKM matrix unitarity}}
  \vskip 10 true mm

  \vskip -2 true mm

  \centerline{
    Alessandro Conigli$\,^{a,b}$,
    Julien~Frison$\,^{c}$,
    Alejandro S\'aez$\,^{d}$
    }

  \vskip 4 true mm
  \vskip 1 true mm
  \centerline{\it $^a$Helmholtz Institute Mainz, Johannes Gutenberg University, Mainz, Germany}
  \vskip 1 true mm
  \centerline{\it $^b$GSI Helmholtz Centre for Heavy Ion Research, Darmstadt, Germany}
  \vskip 1 true mm
  \centerline{\it $^c$John von Neumann-Institut f\"ur Computing NIC,}
  \centerline{\it Deutsches Elektronen-Synchrotron DESY, Platanenallee 6, 15738 Zeuthen, Germany}
  \vskip 1 true mm
  \centerline{\it $^d$Instituto de F\'isica Corpuscular (IFIC), CSIC-Universitat de Val\`encia, 46071, Valencia, Spain}
  \vskip 15 true mm


  %
  \noindent{\bf Abstract:}
  We present lattice results for $f_K/f_{\pi}$ in the iso-symmetric limit of pure QCD (isoQCD) with $N_f=2+1$ flavours, along with a determination of $|V_{us}|/|V_{ud}|$ and a study on the unitarity of the first row of the Cabibbo-Kobayashi-Maskawa (CKM) matrix after introducing strong isospin-breaking and QED effects.
  The results obtained are based on a combination of a Wilson unitary action and the mixed-action setup introduced in \cite{Bussone:2025wlf,Bussone:2023kag}.
  The combination of the two regularisations enables a more precise control over the continuum-limit extrapolation.
  \vspace{10truemm}
  %


\section{Introduction}

In precision era hadronic physics, the ability to make precise predictions of the Standard Model is of the utmost importance for conducting both direct and indirect searches for New Physics.
The quark-flavour sector of the Standard Model constitutes a rich arena for this undertaking, where the non-perturbative nature of the strong coupling plays a crucial role.
In this context, a first-principles approach to compute hadronic contributions to different physical phenomena, as provided by Lattice QCD, is required.
Examples of such processes where non-perturbative input is needed include leptonic and semileptonic meson decays, which provide access to the Cabibbo-Kobayashi-Maskawa (CKM) matrix elements.
The main topic of this work is the computation of $f_{K^{\pm}}/f_{\pi^{\pm}}$.
This quantity allows to extract the ratio of CKM matrix elements $|V_{us}|/|V_{ud}|$ employing the experimental photon inclusive ratio of decay rates
\begin{equation}
  \label{eq:branching}
\frac{\Gamma\left(K\rightarrow l\overline{\nu}_l[\gamma]\right)}{\Gamma\left(\pi\rightarrow l\overline{\nu}_l[\gamma]\right)}=\frac{|V_{us}|^2}{|V_{ud}|^2}\frac{f_{K^{\pm}}^2}{f_{\pi^{\pm}}^2}\frac{m_{K^{\pm}}\left(1-\frac{m_l^2}{m_{K^{\pm}}^2}\right)^2}{m_{\pi^{\pm}}\left(1-\frac{m_l^2}{m_{\pi^{\pm}}^2}\right)^2}\times\left(1+\delta_{\rm EM}^K-\delta_{\rm EM}^{\pi}\right),
\end{equation}
where $\delta_{\rm EM}^{K(\pi)}$ parameterises the QED contributions to $\Gamma\left(K(\pi)\rightarrow l\overline{\nu}_l[\gamma])\right)$  \cite{Marciano:1993sh,Cirigliano:2011tm}.
Furthermore, one can subsequently obtain a determination of $|V_{us}|$ employing the value of $|V_{ud}|$ as given from super-allowed nuclear $\beta$-decays \cite{ParticleDataGroup:2024cfk}, and perform a unitarity test of the first row of the CKM matrix.
In the lattice, the isospin-symmetric limit of pure QCD (isoQCD) is what is most often simulated, with two mass-degenerate light quarks and vanishing electromagnetic coupling $\alpha=0$.
In order to gain access to the ratio $f_{K^{\pm}}/f_{\pi^{\pm}}$ in the full theory, we compute $f_K/f_{\pi}$ in isoQCD and later correct for strong isospin-breaking effects.
For these effects, as well as for the QED corrections in \cref{eq:branching}, different approaches exist in the literature \cite{Marciano:1993sh,Cirigliano:2011tm,DiCarlo:2019thl,Boyle:2022lsi}.
Though in principle it is desirable to have a fully non-perturbative determination of these corrections, we find that the final error of our results is completely dominated by the lattice uncertainty of $f_K/f_{\pi}$ in isoQCD, reducing thus the impact of what one uses to correct for strong isospin-breaking and QED effects.
The computation of the ratio $f_K/f_{\pi}$ in isoQCD is an unambiguous prediction that can be directly compared to lattice computations by other groups, as long as the input needed to define the physical point is fixed, e.g. as in the Edinburgh Consensus \cite{FlavourLatticeAveragingGroupFLAG:2024oxs}.
For our lattice setup, we employ $N_f=2+1$ flavours, i.e. two mass-degenerate light and one strange flavours, based on CLS ensembles \cite{Bruno:2014jqa,Mohler:2017wnb}.
Two different lattice regularisations are employed: the first is a Wilson unitary one \cite{Bruno:2014jqa}, with Wilson fermions both in the sea and the valence sectors; the second is a mixed action one \cite{Bussone:2025wlf,Bussone:2023kag}, employing Wilson fermions in the sea and Wilson twisted mass (Wtm) \cite{Frezzotti:2000nk,Pena:2004gb} fermions in the valence.
The latter was properly tuned in order to guarantee unitarity in the continuum limit by matching the pion and kaon masses in the sea and valence sectors.
This ensures that the physical quark masses in both sectors agree, recovering unitarity in the continuum limit.
The combination of both lattice regularisations enables a precise control of the continuum-limit extrapolation and the enhancement of the statistical precision of the final result.
All the ensembles employed have a volume of $m_{\pi}L\gtrsim 4$ and we correct for finite-volume effects through (finite-volume) next-to-leading order (NLO)  Chiral Perturbation Theory ($\chi$PT) for all the observables of interest.
We find the correction to be by far subleading at the current statistical precision (cf. \res{sec:fve}).
In order to define the physical point at which to perform the chiral interpolation of our $f_K/f_{\pi}$ lattice data we employ the pion and kaon pseudoscalar masses, expressed in units of the pion decay constant.
This is equivalent to setting the scale with $f_{\pi}$.
These three quantities are prescribed by the Edinburgh Consensus\footnote{\url{https://indico.ph.ed.ac.uk/event/257/}} and \mbox{FLAG 24}
\cite{FlavourLatticeAveragingGroupFLAG:2024oxs} to be
\begin{align}
  \label{eq:Edinburgh}
  m_{\pi}^{\rm ph}&=135\;{\rm MeV}, \quad m_K^{\rm ph}=494.6\;{\rm MeV}, \quad f_{\pi}^{\rm ph}=130.5\;{\rm MeV},
\end{align}
in isoQCD.
Using $f_{\pi}$ to set the scale, as opposed to employing a theory scale (e.g. gradient flow scales $t_0, w_0$ \cite{Luscher:2010iy,BMW:2012hcm}), offers the benefit that its physical value is already known, as evidenced by \cref{eq:Edinburgh}, and thus one does not rely on the continuum-limit extrapolation of one of these scales.
Conversely, $f_{\pi}$ has a strong chiral dependence.
However, this effect can be properly addressed in our analysis thanks to the inclusion of physical point ensembles.
This work is structured as follows.
In \res{sec:setup} we briefly review the basic details of our lattice setup, which is discussed in greater detail in \cite{Bussone:2025wlf,Bussone:2023kag}.
In \res{sec:fve} we introduce our strategy to correct for finite-volume effects in the relevant mesonic observables, based on NLO $\chi$PT.
In \res{sec:chiral-continuum} the strategy for interpolating our $f_K/f_{\pi}$ lattice data to the physical pion and kaon masses is presented, together with the continuum-limit extrapolation.
To this end, we explore different fit functions and cuts in data in order to assess the systematic uncertainty related to the chiral-continuum extrapolation.
In \res{sec:CKM} we introduce strong isospin-breaking effects in the ratio $f_K/f_{\pi}$ and review the impact of QED corrections.
We rely on $\chi$PT results with a conservative uncertainty to account for these effects, finding them compatible with the most recent ab initio lattice computations.
This allows us to extract $|V_{us}|/|V_{ud}|$ and subsequently $|V_{us}|$ by employing a determination of $|V_{ud}|$ from super-allowed $\beta$-decays \cite{ParticleDataGroup:2024cfk}, and to perform a test of the unitarity of the first row of the CKM matrix.
Finally, we present our conclusions in \res{sec:conclusions}.
In Appendix \ref{app_input} we provide supplementary material on the impact of modifying the definition of the isoQCD values of the input quantities $m_{\pi,K}$ and $f_{\pi}$.
We also estimate the derivatives of $f_K/f_{\pi}$ in the isoQCD limit with respect to these quantities.
The estimation of statistical uncertainties has been performed using the $\Gamma$-method \cite{Wolff:2003sm} implementation in the \verb|ADerrors| package \cite{Ramos:2018vgu}.

\section{Setup}
\label{sec:setup}

\begin{figure}[ht!]
  \centering
    \includegraphics[width=.52\textwidth]{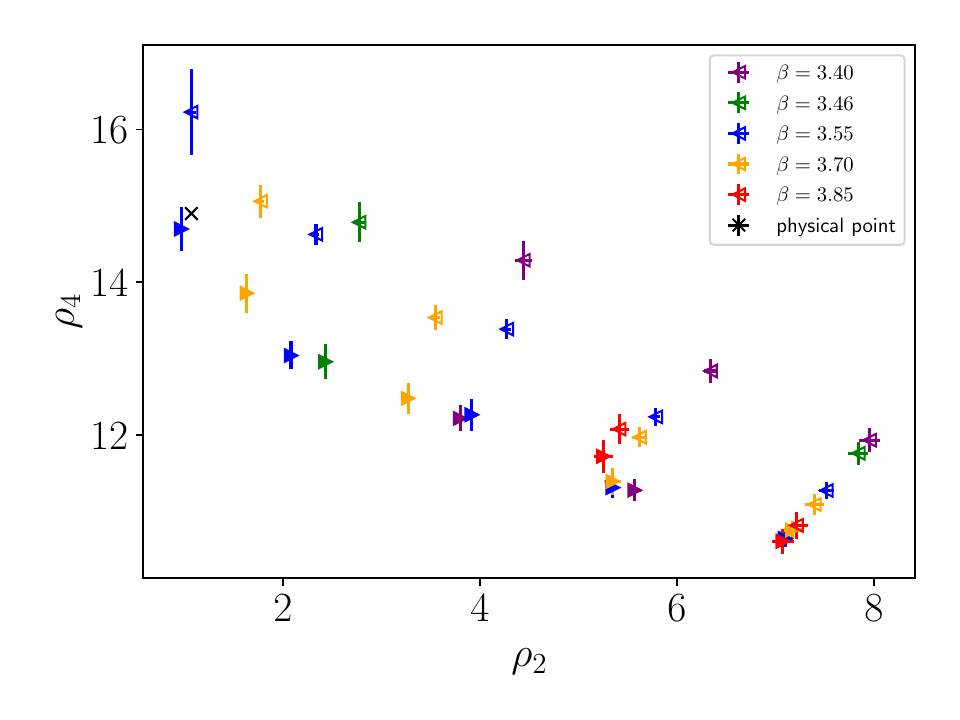}
  \caption{\label{fig:ens} Measured values of $\rho_{2},\;\rho_{4}$, defined in \cref{eq:rho2} and \cref{eq:rho4}, for the set of CLS ensembles employed in this work (cf. \ret{tab:ens}),
    following the ${\rm tr}(M_{q})=2m_{l}+m_{s}\approx{\rm const.}$ chiral
    trajectory.
  Empty points correspond to the Wilson unitary setup, while filled ones correspond to the Wtm mixed action.
  The difference in $\rho_2$ and $\rho_4$ between the Wilson and Wtm mixed action for any given ensemble is due to different cutoff effects in $f_{\pi}$ for each regularisation, as the pion and kaon masses -- in lattice units -- are matched to be the same in both.
  The physical point, marked by a black cross, is reached by interpolating both in $\rho_2$ and $\rho_4$.}
\end{figure}

In this section we review the basic features of our setup, with an emphasis on the aspects that are most relevant to the present work.
We refer the reader to \cite{Bussone:2025wlf,Bussone:2023kag} for a fully detailed discussion of our approach.
The set of lattice ensembles under study were generated by the Coordinated Lattice Simulations (CLS) initiative \cite{Bruno:2016plf,Mohler:2017wnb}.
They use the Lüscher-Weisz gauge action \cite{Kuberski:2023zky} and non-perturbatively $\mbox{O}(a)$-improved $N_f=2+1$ Wilson fermions in the sea sector.
Open boundary conditions (OBC) are employed in the time direction for the gauge degrees of freedom for most ensembles, in order to avoid topology freezing as the continuum limit is approached.
The ensembles that we include in the analysis follow a chiral trajectory defined by an approximately constant trace of the bare quark mass matrix, imposed by the condition
\begin{equation}
  \label{eq:Tr_Mq_const}
8t_0\left(m_K^2+\frac{1}{2}m_{\pi}^2\right)\approx{\rm const.}\Rightarrow{\rm tr}\left(M_q\right)=2m_{l}+m_{s}\approx{\rm const.}
\end{equation}
More details on the ensembles are provided in \ret{tab:ens}.
For our kinematic variables, we define
\begin{align}
  \label{eq:rho2}
  \rho_2&=\frac{m_{\pi}^{2}}{f_{\pi}^{2}}, \\
  \label{eq:rho4}
  \rho_{4}&=\frac{m_{K}^2+\frac{1}{2}m_{\pi}^{2}}{f_{\pi}^{2}}.
\end{align}
With the input in \cref{eq:Edinburgh}, the physical values $\rho_2^{\rm ph},\;\rho_4^{\rm ph}$ are unambiguously determined as
\begin{align}
  \label{eq:ph_point}
  \rho_2^{\rm ph}=1.070, \quad \rho_{4}^{\rm ph}=14.899,
\end{align}
thereby defining the physical point at which to chirally-interpolate the $f_K/f_{\pi}$ data.
Note that we interpolate both in $\rho_2$ and $\rho_4$.
We have one ensemble (E250) lying at the physical pion mass.
In \refig{fig:ens} we show the range of values explored for these quantities along the ensembles considered in this work.
In the valence sector, the regularisation employed for the Dirac operator is either identical to that utilised in the sea sector (Wilson unitary setup), or a Wilson twisted mass \cite{Frezzotti:2000nk,Pena:2004gb} one (mixed-action setup).
In the latter case, the valence parameters were tuned in order to match the pion and kaon masses -- in units of the lattice spacing -- in the sea and valence sectors \cite{Bussone:2025wlf,Bussone:2023kag}.
This ensures that physical quark masses coincide in both sectors, thereby recovering unitarity of the theory in the continuum limit, as was checked in \cite{Bussone:2025wlf} through universality tests.
Furthermore, the valence sector was tuned to maximal twist by setting the valence light PCAC standard quark mass to zero.
This results in automatic $\mbox{O}(a)$-improvement up to residual sea cutoff effects of $\mbox{O}\left(ag_0^4{\rm tr}M_q^{\rm sea}\right)$.
For a more thorough exposition about the tuning of the mixed action, see \cite{Bussone:2025wlf,Bussone:2023kag}.
The two lattice regularisations must share the same continuum limit; however they can exhibit different cutoff effects.
Consequently, the analysis of both sets of data in conjunction with the imposition of a common continuum limit can facilitate the control of the extrapolation to vanishing lattice spacing.
The relevant lattice observables for this study are the pseudoscalar masses $m_{\rm PS}$, employed to define the physical point, and decay constants $f_{\rm PS}$, with ${\rm PS}=\pi,K$.
Both set of observables can be extracted from two-point functions involving the pseudoscalar density and axial current.
They exhibit clear plateaus, which we extract by model averaging different plateau ranges as proposed in \cite{Neil:2023pgt}.
For more details on this latter subject, we refer to \cite{Bussone:2025wlf,Bussone:2023kag}.
In \ret{tab:measurements} we present  measurements for $f_K/f_{\pi},\;\rho_2,\;\rho_4$, while in \ret{tab:fpi_sym} the values of $af_{\pi}$ at the SU(3) symmetric point, defined by $m_{\pi}=m_K\approx420$ MeV, are shown.
The latter will be used as probes for cutoff effects (cf. \res{sec:chiral-continuum}).
In both tables finite-volume effects are already taken into account (cf. \res{sec:fve}).
All physical observables are free of $\mbox{O}(a)$ cutoff effects, either thanks to the automatic $\mbox{O}(a)$-improvement feature of our mixed action, or to the inclusion of the relevant improvement coefficients for the Wilson unitary setup.
Once again we refer to \cite{Bussone:2025wlf,Bussone:2023kag} for details on improvement, as well as for the relevant renormalisation factors for the decay constants.
\begin{table}[H]
  \begin{center}
  \begin{tabular}{c c @{\hspace{2em}} c c c c c c c c c}
    \toprule
    $\beta$ & \makecell{$a$ \\ $[\rm fm]$} & id & $L/a$ & $T/a$ & \makecell{$m_{\pi}$ \\ $[\text{MeV}]$} & \makecell{$m_{K}$ \\ $[\text{MeV}]$} & $m_{\pi}L$ & LMD & MDU & $N_{\rm cnfg}$ \\
    \toprule
    3.40 & 0.085 & H101r000 & 32 & 96 & 426 & 426 & 5.8 & no & 4004 & 1001 \\
    & & H101r001 & 32 & 96 & 426 & 425 & 5.8 & no & 4036 & 1009 \\
    \cline{3-11}
    & & H102r001 & 32 & 96 & 360 & 446 & 4.9 & yes & 3988 & 997 \\
    \cline{3-11}
    & & H102r002 & 32 & 96 & 360 & 446 & 4.9 & yes & 4032 & 1008 \\
    \cline{3-11}
    & & H105r001 & 32 & 96 & 286 & 470 & 3.9 & yes & 3788 & 947 \\
    & & H105r002 & 32 & 96 & 286 & 470 & 3.9 & yes & 4168 & 1042 \\
    \cline{3-11}
    & & H105r005 & 32 & 96 & 286 & 470 & 3.9 & no & 3348 & 837 \\
    \midrule
    3.46 & 0.075 & H400r001 & 32 & 96 & 429 & 429 & 5.2 & no & 2020 & 505 \\
    &  & H400r002 & 32 & 96 & 429 & 429 & 5.2 & no & 2160 & 540 \\
    \cline{3-11}
    &  & D450r011 & 64 & 128 & 221 & 483 & 5.4 & yes & 4000 & 250 \\
    \midrule
    3.55 & 0.063 & N202r001 & 48 & 128 & 418 & 418 & 6.5 & no & 3596 & 899 \\
    \cline{3-11}
    & & N203r000 & 48 & 128 & 350 & 448 & 5.4 & yes & 3024 & 756 \\
    & & N203r001 & 48 & 128 & 350 & 448 & 5.4 & yes & 3148 & 787 \\
    \cline{3-11}
    & & N200r000 & 48 & 128 & 288 & 470 & 4.4 & yes & 3424 & 856 \\
    & & N200r001 & 48 & 128 & 288 & 470 & 4.4 & yes & 3424 & 856 \\
    \cline{3-11}
    & & D200r000 & 64 & 128 & 204 & 488 & 4.2 & yes & 8004 & 2001 \\
    \cline{3-11}
    & & E250r001 & 96 & 192 & 131 & 497 & 4.0 & yes & 4000 & 100 \\
    \midrule
    3.70 & 0.049 & N300r002 & 48 & 128 & 427 & 427 & 5.1 & no & 6084 & 1521 \\
    \cline{3-11}
    & & N302r001 & 48 & 128 & 350 & 457 & 4.2 & no & 8804 & 2201 \\
    \cline{3-11}
    & & J303r003 & 64 & 192 & 261 & 481 & 4.1 & yes & 8584 & 1073 \\
    \cline{3-11}
    & & E300r001 & 96 & 192 & 177 & 499 & 4.2 & no & 4540 & 227 \\
    \midrule
    3.85 & 0.038 & J500r004 & 64 & 192 & 418 & 418 & 5.2 & no & 6288 & 198 \\
    & & J500r005 & 64 & 192 & 418 & 418 & 5.2 & no & 5232 & 130 \\
    & & J500r006 & 64 & 192 & 418 & 418 & 5.2 & no & 3096 & 94 \\
    \cline{3-11}
    & & J501r001 & 64 & 192 & 339 & 453 & 4.3 & no & 6536 & 371 \\
    & & J501r002 & 64 & 192 & 339 & 453 & 4.3 & no & 4560 & 248 \\
    & & J501r003 & 64 & 192 & 339 & 453 & 4.3 & no & 4296 & 168 \\
    \bottomrule
    \end{tabular}
    \end{center}
    \caption{\label{tab:ens}
    Set of $N_f=2+1$ CLS ensembles employed in this work.
    We quote the values of the bare coupling $\beta = 6/g_0^2$; approximate values of the lattice spacing $a$ as determined in \cite{Bussone:2025wlf}; the spatial and temporal lattice extents; the masses of the pion
$m_{\pi}$ and the kaon $m_K$; and $m_{\pi} L$.
    Additionally, the number of configurations $N_{\rm cnfg}$ and the
corresponding length of the Monte Carlo chain in Molecular Dynamics
Units (MDU) employed in the computation of the mesonic correlation
functions are provided.
    Open boundary conditions in the time direction are employed for the gauge fields, with the exception of ensembles E250 and D450, which have anti-periodic boundary conditions in time.
    The column ``LMD'' refers to whether the reweighting factors were computed following \cite{Kuberski:2023zky} (``yes'') or \cite{Luscher:2012av,Kennedy:1998cu,Clark:2003na,Clark:2006fx,Luscher:2019openQCD} (``no'').
    The ensemble id consists of the id used for the ensemble plus a sufix (r000,...) to distinguish different replica runs (independent Monte-Carlo runs which share the same physical and algorithmic parameters, but use different random numbers \cite{Wolff:2003sm}).
    We separate by horizontal lines different ensembles.
    In the case of ensemble H105, we have the two replica H105r001 and H105r002.
    H105r005 cannot be considered a replica of H105 since different algorithmic parameters were used, and therefore it is treated as a different ensemble with the same physical parameters.
    The same applies to H102r001 and H102r002.
    }
  \end{table}
%

%
\begin{longtable}{c c c c c c c}
\toprule
ID & $\rho_2$ [W] & $\rho_2$ [Wtm] & $\rho_4$ [W] & $\rho_4$ [Wtm] & $f_K/f_{\pi}$ [W] & $f_K/f_{\pi}$ [Wtm] \\
\toprule
H101 & $8.316(76)$ & $7.235(52)$ & $12.47(11)$ & $10.85(8)$ & $1.0$ & $1.0$  \\
H102 & $6.542(56)$ & $5.716(45)$ & $13.35(11)$ & $11.66(8)$ & $1.0509(21)$ & $1.0569(16)$  \\
H105 & $4.617(83)$ & $3.873(49)$ & $14.92(23)$ & $12.51(10)$ & $1.1182(78)$ & $1.1021(33)$  \\
\midrule
H400 & $8.474(89)$ & $7.524(77)$ & $12.71(13)$ & $11.29(12)$ & $1.0$ & $1.0$  \\
D450 & $2.859(56)$ & $2.503(32)$ & $15.10(22)$ & $13.22(18)$ & $1.1418(66)$ & $1.1422(33)$  \\
\midrule
N202 & $7.834(60)$ & $7.238(68)$ & $11.75(9)$ & $10.86(10)$ & $1.0$ & $1.0$  \\
N203 & $5.884(38)$ & $5.357(47)$ & $12.61(7)$ & $11.48(10)$ & $1.0540(15)$ & $1.0519(17)$  \\
N200 & $4.351(40)$ & $4.057(36)$ & $13.76(9)$ & $12.83(10)$ & $1.0998(27)$ & $1.1070(34)$  \\
D200 & $2.392(26)$ & $2.142(27)$ & $14.94(11)$ & $13.37(12)$ & $1.1551(33)$ & $1.1343(46)$  \\
E250 & $1.100(43)$ & $0.997(18)$ & $16.41(56)$ & $14.88(26)$ & $1.1952(215)$ & $1.1937(83)$  \\
\midrule
N300 & $7.965(53)$ & $7.635(60)$ & $11.95(8)$ & $11.45(9)$ & $1.0$ & $1.0$  \\
N302 & $5.668(84)$ & $5.427(72)$ & $12.53(16)$ & $12.00(15)$ & $1.0574(20)$ & $1.0608(42)$  \\
J303 & $3.545(35)$ & $3.326(46)$ & $13.97(12)$ & $13.11(16)$ & $1.1188(44)$ & $1.1161(64)$  \\
E300 & $1.889(38)$ & $1.775(29)$ & $15.79(27)$ & $14.83(18)$ & $1.1735(138)$ & $1.1893(201)$  \\
\midrule
J500 & $7.443(130)$ & $7.239(165)$ & $11.16(20)$ & $10.86(25)$ & $1.0$ & $1.0$  \\
J501 & $5.374(88)$ & $5.292(118)$ & $12.30(19)$ & $12.12(26)$ & $1.0607(28)$ & $1.0647(51)$  \\
\bottomrule
\caption{Measurements of $\rho_2,\;\rho_4$ and $f_K/f_{\pi}$ for both the Wilson unitary [W] and Wtm mixed-action [Wtm] setups.
The mixed action was tuned in order to impose identical pion and kaon masses in the sea and valence sectors in units of the lattice spacing.
The difference between $\rho_2,\;\rho_4$ in the Wilson unitary and mixed-action regularisations therefore comes exclusively from different cutoff effects in $f_{\pi}$ for each regularisation.
Finally, the ratio $f_K/f_{\pi}$ is $1.0$ without error in the SU(3) symmetric point by construction.
All measurements are corrected for finite-volume effects as detailed in \res{sec:fve}.
Measurements corresponding to ensembles H105 and H105r005 were averaged together, as well as those of H102r001 and H102r002 (cf. \ret{tab:ens}).
All quantities are renormalised and non-perturbatively $\mbox{O}(a)$-improved, for a general discussion we refer to \cite{Bussone:2025wlf,Bussone:2023kag}.
}
\label{tab:measurements}
\end{longtable}

\begin{longtable}{c c c}
\toprule
$\beta$ & $af_{\pi}^{\rm sym}$ [W] & $af_{\pi}^{\rm sym}$ [Wtm] \\
\toprule
$3.40$ & $0.06319(28)$ & $0.06774(24)$ \\
$3.46$ & $0.05620(25)$ & $0.05965(30)$ \\
$3.55$ & $0.04792(17)$ & $0.04985(19)$ \\
$3.70$ & $0.03770(11)$ & $0.03850(13)$ \\
$3.85$ & $0.02991(23)$ & $0.03033(33)$ \\
\bottomrule
\caption{Values for $af_{\pi}$ at the SU(3) symmetric point, defined by $m_{\pi}=m_K\approx420$ MeV, for each $\beta$ considered in this work, to be employed as probes for the cutoff effects (cf. \res{sec:chiral-continuum}).
We quote values for both the Wilson unitary [W] and Wtm mixed-action [Wtm] setups, the difference between them being a cutoff effect.
Finite-volume effects are included as detailed in \res{sec:fve}.
Results are renormalised and non-perturbatively $\mbox{O}(a)$-improved, for a general discussion we refer to \cite{Bussone:2025wlf,Bussone:2023kag}
}
\label{tab:fpi_sym}
\end{longtable}
%

\section{Finite volume effects}
\label{sec:fve}

\begin{figure}
  \centering
  \includegraphics[width=.49\textwidth]{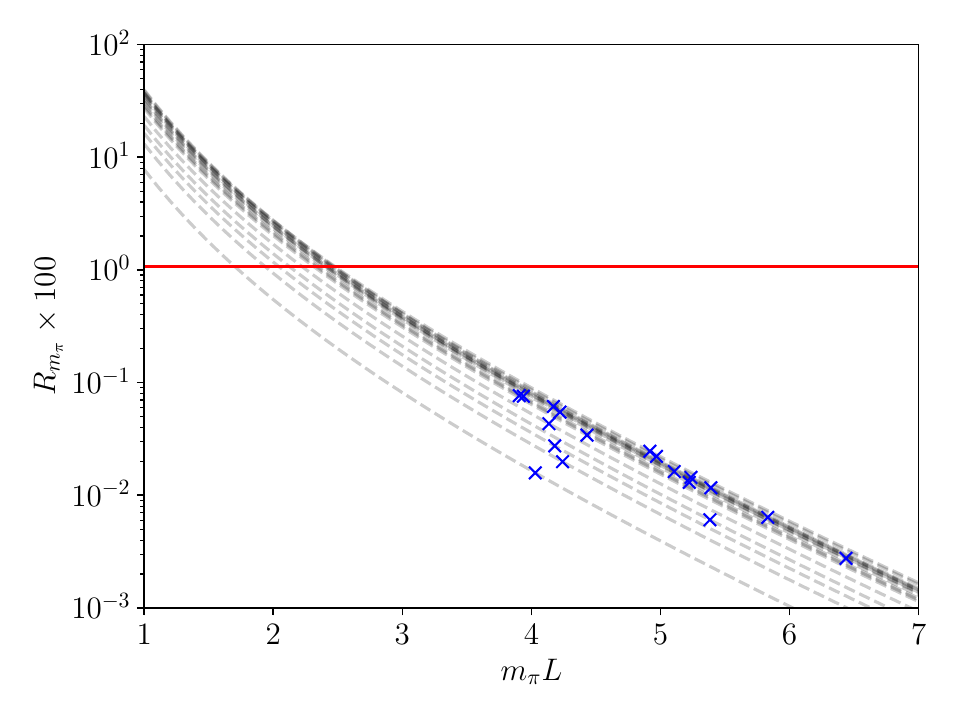}
  \includegraphics[width=.49\textwidth]{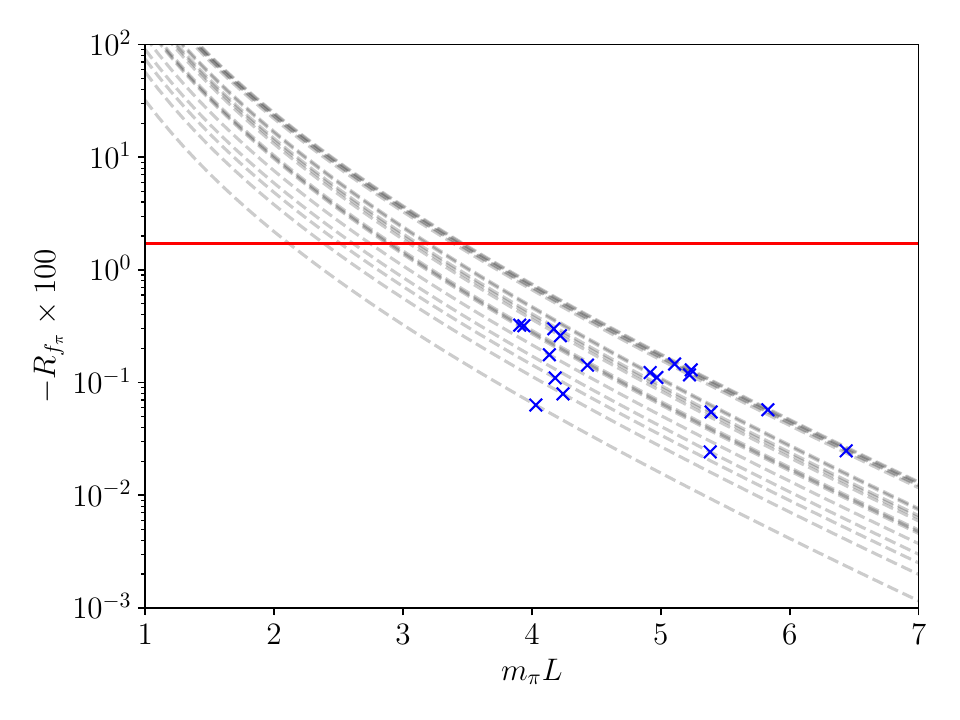}
  \includegraphics[width=.49\textwidth]{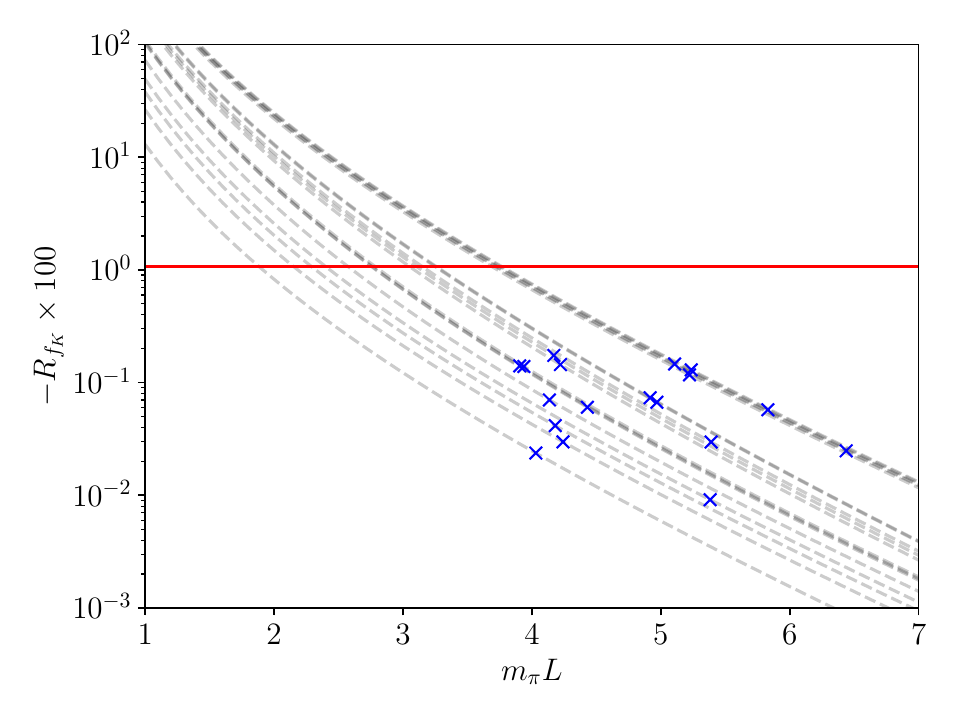}
    \includegraphics[width=.49\textwidth]{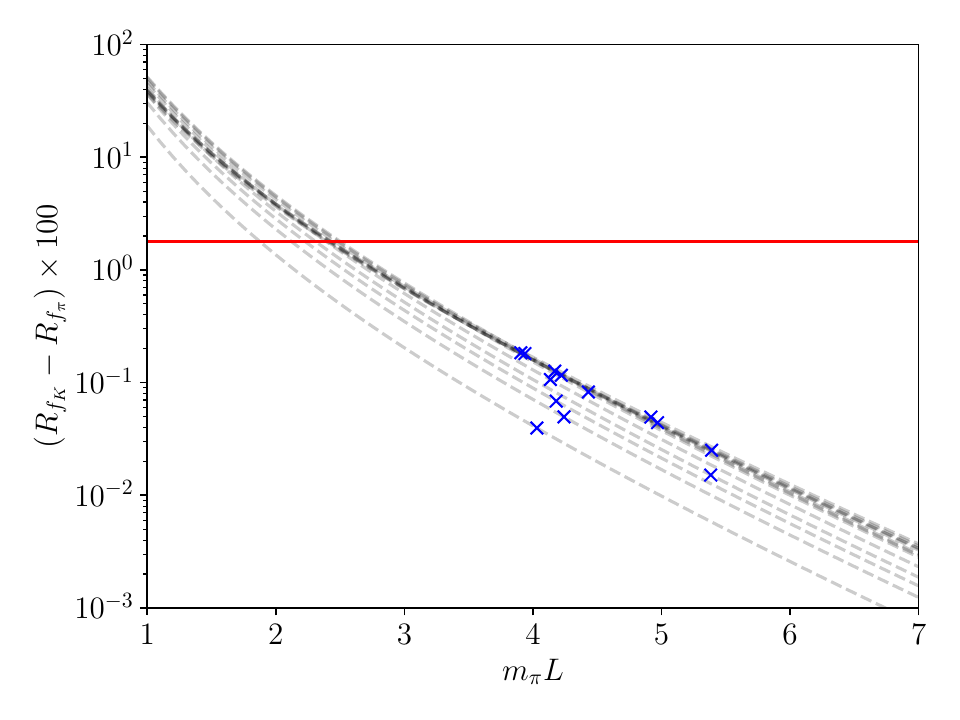}
  \caption{\label{fig:FVE}
   $R_X$ for $X=m_{\pi},f_{\pi},f_K$ for the set of ensembles considered in this work in the Wilson unitary setup, as given by \cref{eq:FVE} (blue crosses), together with the expected dependence on $m_{\pi}L$ for each of these ensembles (grey dashed lines).
  It can be seen that the finite-volume correction is of a few per mille at most, well below the statistical precision of the lattice data (the red horizontal line shows a representative example of the relative statistical precision).
  $R_{f_K}-R_{f_{\pi}}$ in the bottom right plot is the relevant correction factor for the ratio $f_K/f_{\pi}$.}
\end{figure}

The masses and decay constants of the pion and kaon suffer from finite volume
effects, which will be addressed in this section.
Finite-volume $\chi$PT enables to correct these effects in our quantities of interest.
To NLO and for $X=m_{\pi,K},\;f_{\pi,K}$, denoting by $X^{(\infty)}$ the infinite volume quantity and by $X^{(L)}$ the finite volume one \cite{Colangelo:2005cg,Colangelo:2005gd},
\begin{align}
X^{(\infty)}&=X^{(L)}\frac{1}{1+R_X}, \notag \\
R_{m_{\pi}}&=\frac{1}{4}\xi_{\pi}\tilde{g}_1(\lambda_{\pi})-\frac{1}{12}\xi_{\eta}\tilde{g}_1(\lambda_{\eta}),
           \quad R_{m_K}=\frac{1}{6}\xi_{\eta}\tilde{g}_1(\lambda_{\eta}), \notag \\ R_{f_{\pi}}&=-\frac{3}{8}\xi_{\pi}\tilde{g}_1(\lambda_{\pi})-\frac{3}{4}\xi_{K}\tilde{g}_1(\lambda_{K})-\frac{3}{8}\xi_{\eta}\tilde{g}_1(\lambda_{\eta}), \quad R_{f_K}=-\xi_{\pi}\tilde{g}_1(\lambda_{\pi})-\frac{1}{2}\xi_{K}\tilde{g}_1(\lambda_{K}), \notag \\
         \label{eq:FVE}
\xi_{PS}&=\frac{m_{PS}^2}{(4\pi f_{\pi})^2}, \quad \lambda_{PS}=m_{PS}L, \quad \tilde{g}_1(x)=\sum_{n=1}^{\infty}\frac{4m(n)}{\sqrt{n}x}K_1(\sqrt{n}x),
\end{align}
where $m_{\eta}^2=\frac{4}{3}m_K^2-\frac{1}{3}m_{\pi}^2$, $K_1(x)$ is the
Bessel function of the second kind, and $m(n)$ is given in \cite{Colangelo:2005cg,Colangelo:2005gd}.
The shifts in the values of $m_{\pi,K}$ and $f_{\pi,K}$ due to the finite-volume corrections are well within
our present statistical errors, even for the smallest volume considered, corresponding to $m_{\pi}L=3.9$ in the H105 ensemble, as can be seen in \refig{fig:FVE}.

\section{Chiral-continuum extrapolation}
\label{sec:chiral-continuum}

In order to compute the ratio $f_K/f_{\pi}$ in isoQCD at the physical point, we need to chirally-interpolate our lattice data to $\rho_2^{\rm ph},\;\rho_4^{\rm ph}$ as given in \cref{eq:Edinburgh}, and perform the continuum limit to $a\rightarrow0$.
For the former, we can use $\chi$PT-inspired formulae, as well as Taylor
expansions in the pion and kaon masses.
From SU(3) NLO $\chi$PT we have
\begin{align}
  \label{eq:ChPT_fpi}
f_{\pi}&=f\left[1+\frac{16B_0L_5(\mu)}{f^2}m_l+\frac{16B_0L_4(\mu)}{f^2}(2m_l+m_s)-2\mathcal{L}(m_{\pi}^2,\mu)\right. \left.-\mathcal{L}(m_K^2,\mu)\right], \\
\label{eq:ChPT_fk}
f_K&=f\left[1+\frac{8B_0L_5(\mu)}{f^2}(m_l+m_s)+\frac{16B_0L_4(\mu)}{f^2}(2m_l+m_s)\right. \notag\\ &\hspace{1.1cm}\left.-\frac{3}{4}\mathcal{L}(m_{\pi}^2,\mu)-\frac{3}{2}\mathcal{L}(m_K^2,\mu)-\frac{3}{4}\mathcal{L}(m_{\eta}^2,\mu)\right],
\end{align}
where
\begin{align}
  m_{\pi}^2&=2B_0m_l, \quad m_K^2=B_0(m_l+m_s), \quad m_{\eta}^2=\frac{4}{3}m_K^2-\frac{1}{3}m_{\pi}^2, \\
\mathcal{L}(x,\mu)&=\frac{x}{(4\pi f)^2}\textrm{log}\frac{x}{\mu^{2}},
\end{align}
with $f$ the decay constant in the chiral limit, and $\mu$ a renormalisation scale.
The dependence of the LECs $L_4(\mu),\;L_5(\mu)$ in $\mu$ cancels with that of the chiral logarithms $\mathcal{L}(x,\mu)$, so that physical quantities are $\mu$-independent.
Computing the ratio of decay constants gives
\begin{equation}
  \label{eq:ratio}
\frac{f_K}{f_{\pi}}=\frac{1+8L_5(\mu)q^2\left(\rho_4-\frac{1}{2}\rho_2\right)-\frac{3}{4}\mathcal{L}_{\pi}(\mu)-\frac{3}{2}\mathcal{L}_K(\mu)-\frac{3}{4}\mathcal{L}_{\eta}(\mu)}{1+8L_5(\mu)q^2\rho_2-2\mathcal{L}_{\pi}(\mu)-\mathcal{L}_K(\mu)},
\end{equation}
where
\begin{align}
  q&\equiv\frac{f_{\pi}}{f},\\
  \label{eq:Lpi}
  \mathcal{L}_{\pi}(\mu)&=\frac{\rho_{2}}{(4\pi)^2}q^2\log\left(\frac{m_{\pi}^{2}}{\mu^2}\right), \\
  \label{eq:LK}
  \mathcal{L}_{K}(\mu)&=\frac{\rho_{4}-\frac{1}{2}\rho_{2}}{(4\pi )^2}q^2\log\left(\frac{m_K^2}{\mu^2}\right),
  \\
  \label{eq:Leta}
    \mathcal{L}_{\eta}(\mu)&=\frac{\frac{4}{3}\rho_{4}-\rho_{2}}{(4\pi)^2}q^2\log\left(\frac{m_{\eta}^2}{\mu^2}\right).
\end{align}
We choose $\mu\equiv f_{\pi K}=\frac{2}{3}\left(f_K+\frac{1}{2}f_{\pi}\right)$ in the spirit of keeping it a constant renormalisation scale, since $f_{\pi K}$ has a weak chiral dependence along our ${\rm tr}\left(M_q\right)\approx{\rm const.}$ trajectory \cite{Bruno:2016plf}.
Furthermore, we treat the
$q=f_{\pi}/f$ factor as a fit parameter, together with the LEC $L_5$.
This amounts to ignoring the chiral dependence of $q$ due to that of $f_{\pi}$, since it enters as a higher order effect.
We observe no discrepancy with this assumption when performing the chiral fits.
To the expression in \cref{eq:ratio} we need to add cutoff effects, which can be effectively done by expanding the
LECs $f,L_{5}$ as
\begin{align}
  L_5&\rightarrow L_{5}\left(1+\mbox{O}\left(a^2\alpha_{S}^{\Gamma_{i}}\right)\right), \notag \\
\label{eq:LEC_a}
  f&\rightarrow f\left(1+\mbox{O}\left(a^2\alpha_{S}^{\Gamma_{i}}\right)\right),
\end{align}
with $\Gamma_{i}$ the various anomalous dimensions provided for our actions in
\cite{Husung:2022kvi}, together with the case $\Gamma_{i}=0$.
This choice for the form of cutoff effects ensures that $f_K/f_{\pi}$ is exactly $1$ at the SU(3) symmetric limit, even at finite lattice spacing.
As probe for cutoff effects, in the spirit of not relying on any theory scale for the analysis, we will employ $af_{\pi}^{\rm sym}(\beta)$ (cf. \ret{tab:fpi_sym}), the pion decay constant (in lattice units) at the SU(3) symmetric point for each value of the inverse coupling $\beta$.
In practice, we substitute
\begin{align}
  L_5q^2&\rightarrow L_{5}q^2\left(1+C_1\left(af_{\pi}^{\rm
    sym}\right)^2\alpha_{S}^{\Gamma_i}\right), \notag \\
  \label{eq:cutoff_parameters}
  q^2&\rightarrow q^2\left(1+C_2\left(af_{\pi}^{\rm
    sym}\right)^2\alpha_{S}^{\Gamma_i}\right),
\end{align}
in \cref{eq:ratio}, with $C_{1,2}$ additional fit parameters.
In order to explore systematic effects in the chiral-continuum extrapolation, it is necessary to test different fit functions and cuts in the data, which we eventually average together.
To this end, the first possible variation involves exploring the different available values of $\Gamma_i$.
However, we observe no sensitivity in the data to this variation,\footnote{The central value of $f_K/f_{\pi}$ in the continuum changes only a $0.4\%$ of $\sigma$ when varying from $\Gamma_{\rm min}=-0.11$ to $\Gamma_{\rm max}=0.76$, corresponding to the minimum possible and maximum known values of $\Gamma_i$ for both our Wilson unitary and Wtm mixed-action setups \cite{Husung:2022kvi}.} and thus choose to perform the final analysis only with $\Gamma_{i}=0$ in order not to explore redundant models.
We note that the smallest possible value of $\Gamma_{i}$ for both our actions is $\Gamma_i=-0.11$.
The ratio in \cref{eq:ratio} can also be expanded to give a new model
\begin{equation}
  \label{eq:ratio_expanded}
\frac{f_K}{f_{\pi}}=1-12L_5(\mu)q^2\rho_2+8L_5(\mu)q^2\rho_4+\frac{5}{4}\mathcal{L}_{\pi}(\mu)-\frac{1}{2}\mathcal{L}_K(\mu)-\frac{3}{4}\mathcal{L}_{\eta}(\mu),
\end{equation}
where again we use \cref{eq:cutoff_parameters} in order to add cutoff effects.
Another possibility for model exploration is to use $\mu=f_{\pi}$ in the definition of the chiral logarithms in \cref{eq:Lpi,eq:LK,eq:Leta} and the LEC $L_5$.
Again, this assumes that the chiral dependence introduced in $\mu$ can be neglected as a higher order effect.
We can also use a Taylor expansion in $\rho_2$ and $\rho_4$.
In order to constrain that the expansion goes to 1 in the SU(3) symmetric limit, we expand the chiral logarithms $\mathcal{L}_{\pi},\;\mathcal{L}_K,\;\mathcal{L}_{\eta}$ around the symmetric point in \cref{eq:ratio_expanded}, arriving at
\begin{equation}
\frac{f_K}{f_{\pi}}=1+A(2\rho_4-3\rho_2)+B\left(\frac{3}{8}\rho_2^2-\frac{11}{6}\rho_4^2+\frac{5}{2}\rho_2\rho_4\right).
\end{equation}
We find that expanding both $A,\;B$ in an analogous way to \cref{eq:LEC_a} always results in the fit parameter controlling the cutoff effects in $B$ being compatible with zero and thus in inflated errors,\footnote{\label{foot:zero_parm}Furthermore, when employing an information criterion-based model average, e.g. employing the Takeuchi Information Criterion \cite{Frison:2023lwb} or the Akaike Information Criterion \cite{Neil:2023pgt}, this model always results in zero weight because of this behaviour.} so we only consider
\begin{equation}
  \label{eq:Taylor_constrained}
\frac{f_K}{f_{\pi}}=1+\left(A+C\left(af_{\pi}^{\rm sym}\right)^2\alpha_{\rm S}^{\Gamma_i}\right)\times(2\rho_4-3\rho_2)+B\left(\frac{3}{8}\rho_2^2-\frac{11}{6}\rho_4^2+\frac{5}{2}\rho_2\rho_4\right),
\end{equation}

\begin{figure}
  \centering
  \includegraphics[width=.7\textwidth]{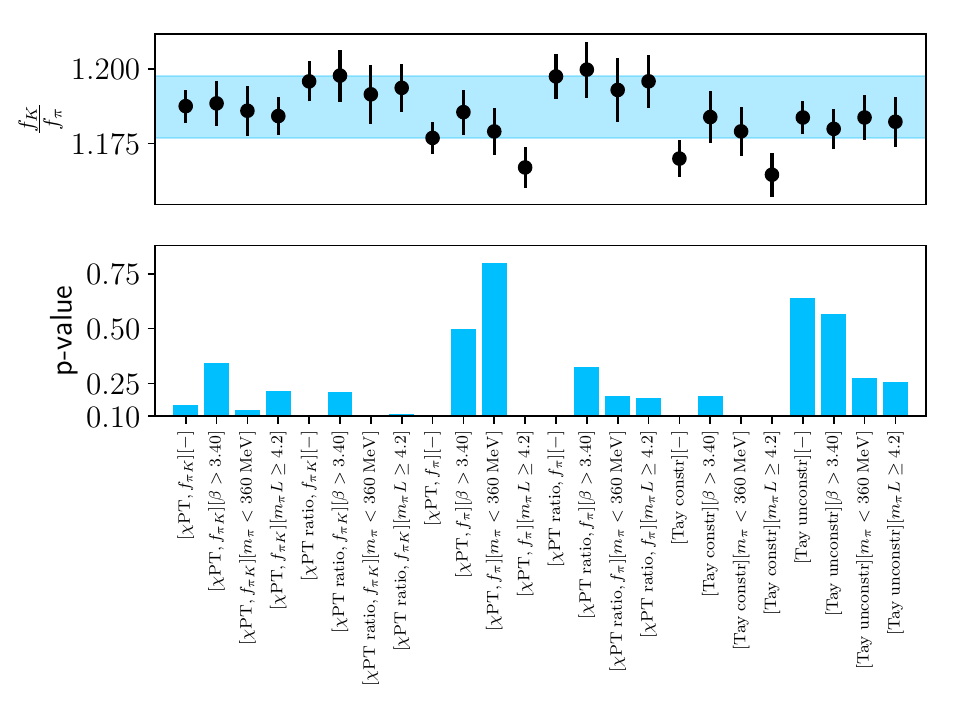}
  \caption{\label{fig:model_av}
  Model exploration for the chiral-continuum extrapolation of the
  ratio $f_K/f_{\pi}$.
In the first bracket, labels $[\chi \rm PT]$ and $[\chi \rm PT\;ratio]$ refer to the use of \cref{eq:ratio_expanded} and \cref{eq:ratio} respectively, with cutoff effects added according to \cref{eq:cutoff_parameters}, while $f_{\pi K}$ or $f_{\pi}$ refers to what was used for the scale $\mu$ inside the chiral logarithms in \cref{eq:Lpi,eq:LK,eq:Leta}.
[Tay constr] refers to \cref{eq:Taylor_constrained}, and [Tay unconstr] to \cref{eq:Taylor_un}.
The second label indicates which data are included: [$-$] means all data
are used, $[\beta>3.40]$ that the coarsest lattice spacing is removed, $[m_{\pi}<360{\rm\;MeV}]$ that only ensembles with pion masses smaller than $360$ MeV are included in the analysis, and
$[m_{\pi}L\geq4.2]$ the same for lattices with the volume satisfying this condition.
The horizontal band in the top panel represents the final quoted
result in \cref{eq:fKfpi_ph}, with statistical and systematic uncertainties added in quadrature.
  }
\end{figure}

\begin{figure}
  \centering
  \includegraphics[width=.5\textwidth]{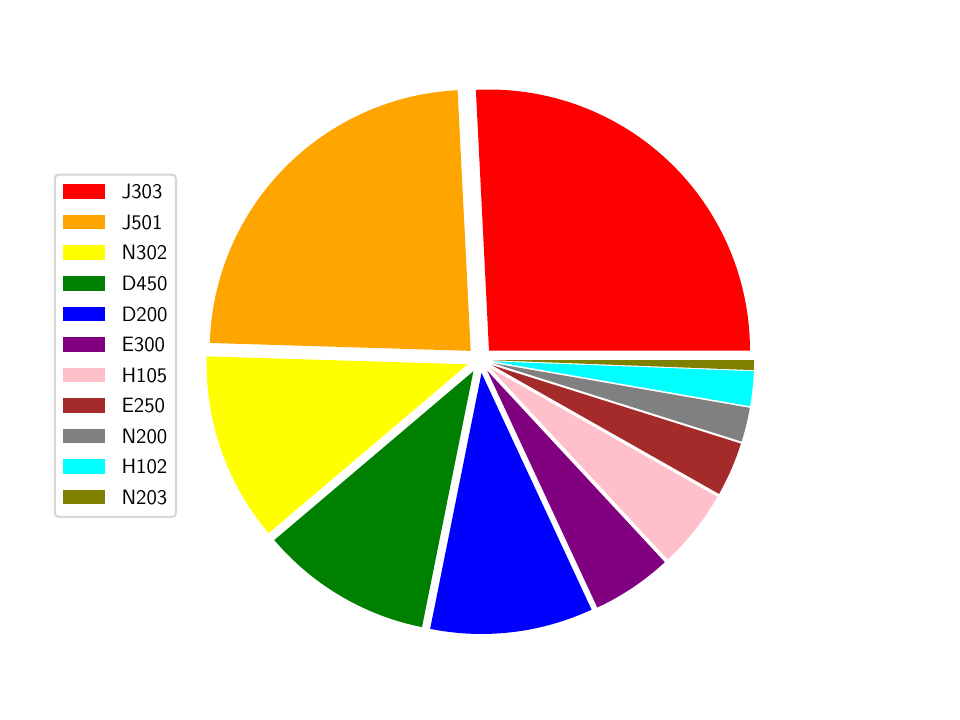}
  \caption{\label{fig:pie} Contribution of the different lattice ensembles to the statistical error squared in \cref{eq:fKfpi_ph}.
  Large contributions indicate that the corresponding ensembles are strongly constraining the chiral-continuum extrapolation of $f_K/f_{\pi}$, and therefore are expected to be ensembles close to the continuum limit and physical point.
  In particular, ensemble J501 is the only ensemble at the finest lattice spacing ($a\approx0.038$ fm) that lies outside the SU(3) symmetric point.
  Ensemble J303, on the other hand, is at the second finest lattice spacing ($a\approx0.049$ fm) and close to the physical point ($m_{\pi}\approx261$ MeV).
  Although we also have an ensemble at the same lattice spacing that is closer to the physical point (E300, $m_{\pi}\approx177$ MeV), its statistical sample is significantly smaller (cf. \ret{tab:ens} and \ret{tab:measurements}), which reduces its overall impact on the fits.
  }
\end{figure}

\begin{figure}
  \centering
  \includegraphics[width=.75\textwidth]{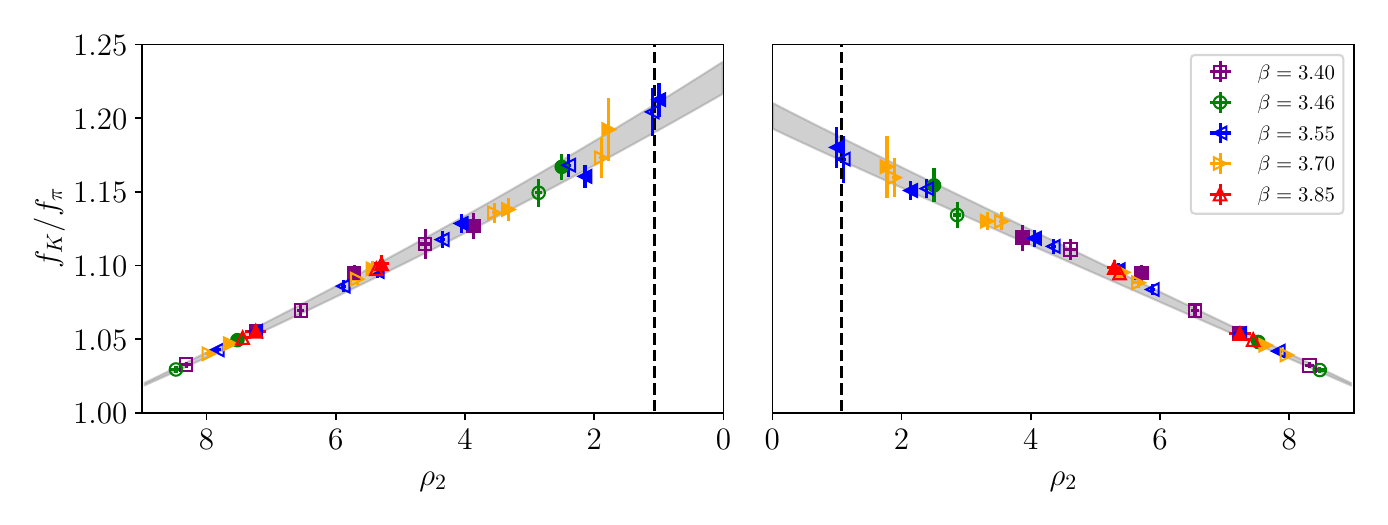}
  \includegraphics[width=.75\textwidth]{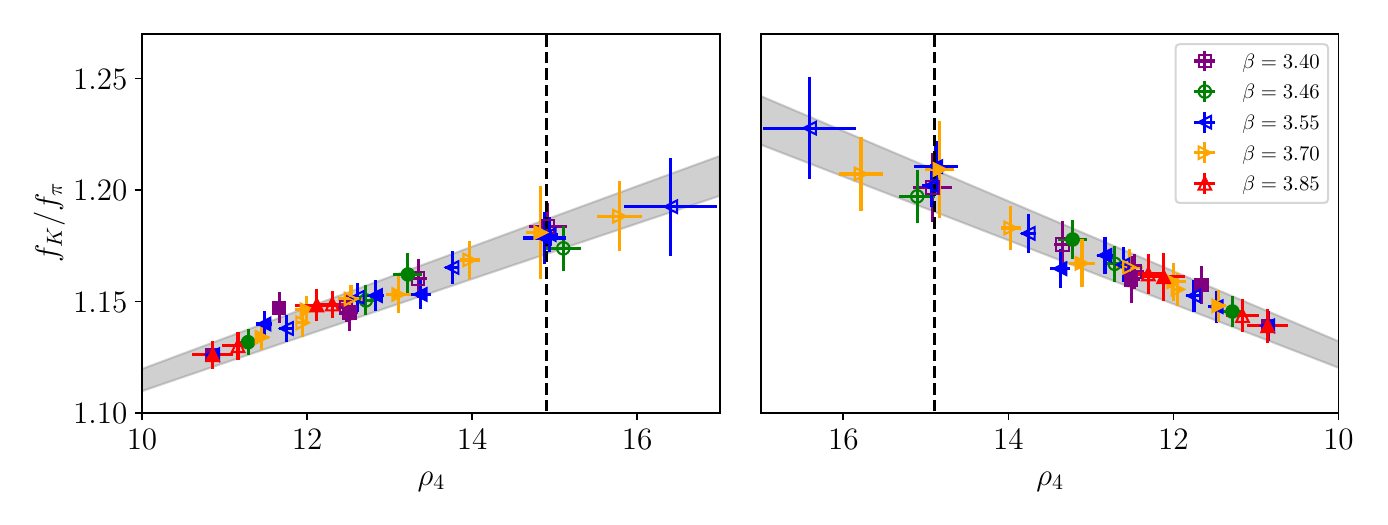}
    \includegraphics[width=.75\textwidth]{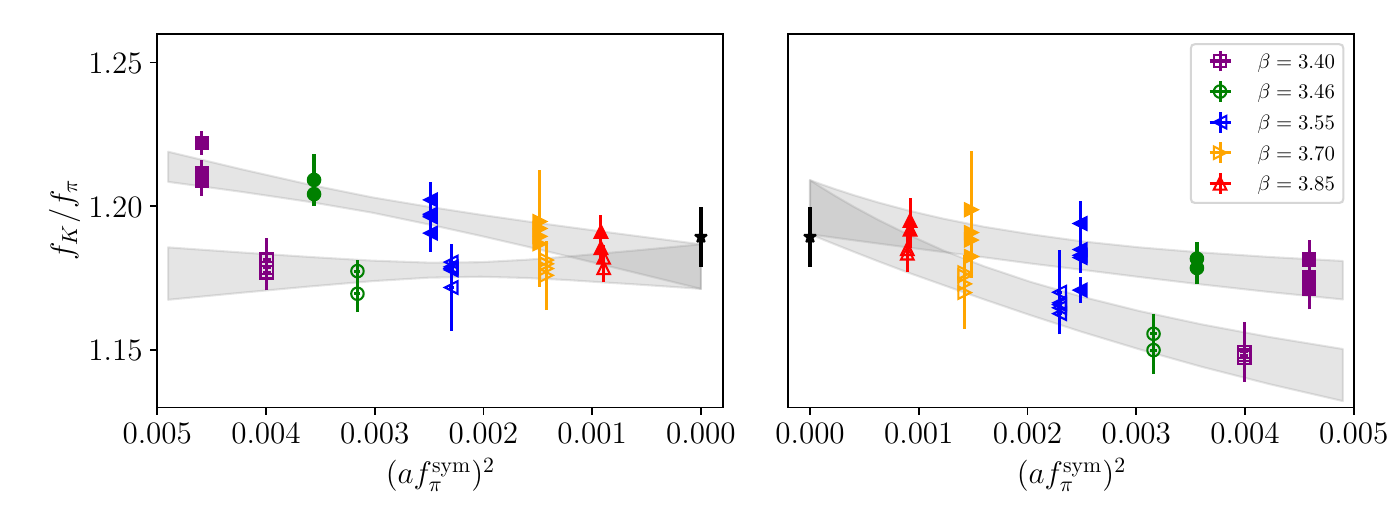}
    \caption{\label{fig:chiral-continuum}
      \textit{Left column}: chiral-continuum extrapolation employing \cref{eq:ratio_expanded}, using $\mu=f_{\pi}$ in the definition of the chiral logarithms, and adding cutoff effects according to \cref{eq:cutoff_parameters} with
      $\Gamma_{i}=0$, removing $m_{\pi}>360$ MeV.
      We note that the $x$-axis direction has been reversed relative to the one in the right column in order to facilitate a direct comparison of the physical point extrapolation in both fits.
      \textit{Right column}: chiral-continuum extrapolation employing \cref{eq:ratio}, using $\mu=f_{\pi}$, and adding cutoff effects according to \cref{eq:cutoff_parameters} with
      $\Gamma_{i}=0$, removing the coarsest lattice spacing $\beta=3.40$.
    In the two top plots, the points in the $y$-axis are projected to $\rho_4^{\rm ph}$ and vanishing lattice spacing, while in the two middle plots they are projected to $\rho_2^{\rm ph}$ and vanishing lattice spacing.
    The vertical dashed line marks the position of the physical point, and the bands correspond to the mass-dependence in the continuum.
    Finally, in the two bottom plots the data in the $y$-axis are projected to both $\rho_2^{\rm ph},\;\rho_4^{\rm ph}$.
    The two bands in each plot correspond to the cutoff dependence for both the Wilson unitary and Wtm mixed-actions setups.
    Empty points are computed with the Wilson unitary action,
    while filled ones are with the mixed action.
The two models shown correspond to the ones giving the lowest and highest results that enter in the model average (cf. \refig{fig:model_av}), and thus define the width of our error band in \cref{eq:fKfpi_ph}, corresponding to the black star point in the two bottom plots.
}
\end{figure}

Finally, we also test a Taylor fit without any constraint in the SU(3) symmetric limit.
In this case we find that the fit is only sensitive to the following terms
\begin{equation}
  \label{eq:Taylor_un}
\frac{f_K}{f_{\pi}}=A_0+A_1\rho_2+A_2\rho_2^2+B_1\rho_4+C\left(af_{\pi}^{\rm sym}\right)^2\alpha_{\rm S}^{\Gamma_i}.
\end{equation}
Adding any other term results in coefficients compatible with zero and a continuum limit compatible with that of the model in \cref{eq:Taylor_un} with inflated errors (cf. footnote \ref{foot:zero_parm}).
Finally, in addition to the different fit function variations given above, we also explore performing cuts in the data.
In particular, we test removing the coarsest lattice spacing $\beta=3.40$, the heaviest pions $m_{\pi}<360$ MeV,\footnote{Note that the symmetric point, corresponding to $m_{\pi}\approx420$ MeV, having by construction $f_K/f_{\pi}=1$ with no error, does not contribute to the $\chi^2$.} and removing volumes with $m_{\pi}L<4.2$.
In \refig{fig:chiral-continuum} we show the chiral-continuum extrapolation
for two representative models, with different fit functions and cuts in the data.
In order to quantify the quality of fits we employ the p-value \cite{Bruno:2022mfy}.
After selecting all models with p-value $>0.1$, we employ the following scheme to compute a model average result:\footnote{No significant change in the final result is seen by performing a model average based on the Takeuchi Information Criterion \cite{Frison:2023lwb} or the Akainke Information Criterion \cite{Neil:2023pgt}.} (i) compute a model average central value and statistical uncertainty with a weighted average of all models, weighting with the errors; (ii) compute a total model average uncertainty (systematic and statistical) by taking half the maximum spread between models; and (iii) considering that systematic and statistical uncertainties add up in quadrature to give the total model average uncertainty, extract the corresponding systematic uncertainty from (i) and (ii).
The result is
\begin{equation}
  \label{eq:fKfpi_ph}
\frac{f_K}{f_{\pi}}=1.1872(59)_{\rm stat}(84)_{\chi-{\rm cont}},
\end{equation}
where ``$\chi-{\rm cont}$'' labels the systematic uncertainty of the model average.
The model average is presented in \refig{fig:model_av}.
We observe the maximum spread between models to be associated with models that employ $\mu=f_{\pi}$ or $\mu=f_{\pi K}$ in \cref{eq:Lpi,eq:LK,eq:Leta}; the use of \cref{eq:ratio_expanded} or \cref{eq:ratio}; and cuts in the pion masses.
We interpret this fact as signaling that higher order effects in the chiral dependence of $f_K/f_{\pi}$ are the main source of systematic uncertainty in \cref{eq:fKfpi_ph}, and thus the need in the future to include NNLO terms in \cref{eq:ChPT_fpi,eq:ChPT_fk}.
We stress that with our current data set we are not able to resolve these terms.
Finally, in \refig{fig:pie} we show the contribution of the different lattice ensembles to the statistical error in \cref{eq:fKfpi_ph}.
%


%
\section{CKM matrix unitarity}
\label{sec:CKM}
\begin{figure}
  \centering
  \includegraphics[width=.6\textwidth]{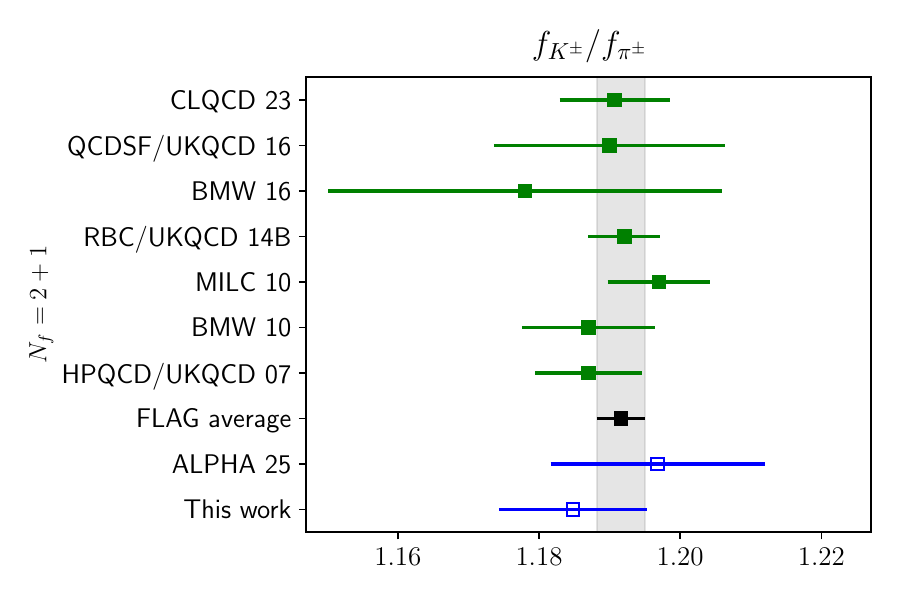}
  \caption{\label{fig:FLAG}
  Comparison of our result for $f_{K^{\pm}}/f_{\pi^{\pm}}$ with strong isospin-breaking corrections, as given in \cref{eq:fKfpi_pureQCD}, to other group's results entering the FLAG average for $N_f=2+1$ \cite{FlavourLatticeAveragingGroupFLAG:2024oxs}.
  CLQCD 23 refers to \cite{CLQCD:2023sdb}, QCDSF/UKQCD 16 to \cite{QCDSF-UKQCD:2016rau}, BMW 16 to \cite{Durr:2016ulb}, RBC/UKQCD 14B to \cite{RBC:2014ntl}, MILC 10 to \cite{MILC:2010hzw}, BMW 10 to \cite{BMW:2010xmi} and HPQCD/UKQCD 07 to \cite{Follana:2007uv}.
  The point ALPHA 25 corresponds to the determination of $f_K$ in \cite{Bussone:2025wlf} based on an analysis setting the scale employing the intermediate scale $\sqrt{t_0}$.}
\end{figure}

In order to test the unitarity of the first row of the CKM matrix
\begin{equation}
  \label{eq:}
|V_{ud}|^2+|V_{us}|^2+|V_{ub}|^2\stackrel{?}{=}1,
\end{equation}
it is necessary to have access to these three matrix elements.
To this end, $|V_{ud}|$ can be extracted with high precision from super-allowed nuclear $\beta$-decays, while $|V_{ub}|$ can be neglected given its small value and the current precision for the other two matrix elements \cite{ParticleDataGroup:2024cfk}.
Lastly, we extract $|V_{us}|/|V_{ud}|$ through the use of \cref{eq:branching}, a lattice determination of $f_{K^{\pm}}/f_{\pi^{\pm}}$ -- after including strong isospin-breaking effects in \cref{eq:fKfpi_ph} -- and taking into account the QED correction factor $\left(1+\delta_{\rm EM}^{K}-\delta_{\rm EM}^{\pi}\right)$.
This allows us to extract $|V_{us}|$ by using the aforementioned value of $|V_{ud}|$.
In order to include strong isospin-breaking effects in our determination of $f_K/f_{\pi}$ given in \cref{eq:fKfpi_ph}, to NLO in SU(3) $\chi$PT \cite{Cirigliano:2007ga,FlavourLatticeAveragingGroupFLAG:2024oxs}
\begin{align}
  \label{eq:rat_corr_master_formula}
\frac{f_{K^{\pm}}^2}{f_{\pi^{\pm}}^2}&=\frac{f_K^2}{f_{\pi}^2}\times\left(1+\delta_{\rm SU(2)}\right), \\
\delta_{\rm SU(2)}&=\sqrt{3}\epsilon_{\rm SU(2)}\left[-\frac{4}{3}\left(f_{K}/f_{\pi}-1\right)+\frac{2}{3(4\pi)^{2}f^2}\left(m_K^2-m_{\pi}^2-m_{\pi}^2\log\frac{m_K^2}{m_{\pi}^2}\right)\right], \\
\epsilon_{\rm SU(2)}&=\sqrt{3}/4R, \quad R\equiv\frac{m_s-m_{ud}}{m_d-m_{u}}.
\end{align}
To compute $\delta_{\rm SU(2)}$, we use
$f=113(28)$ MeV, $R=38.1(1.5)$ as reported by \cite{FlavourLatticeAveragingGroupFLAG:2024oxs}, resulting in
\begin{equation}
  \label{eq:}
\delta_{\rm SU(2)}=-0.0038(7).
\end{equation}
However, given the unknown size of higher order corrections, we opt to take a conservative approach and assign a 100\% uncertainty to $\delta_{\rm SU(2)}$, resulting in
\begin{equation}
  \label{eq:delta_SU2_result}
\delta_{\rm SU(2)}=-0.004(4).
\end{equation}
With this result, we obtain for the ratio of decay constants, including strong isospin-breaking corrections
\begin{equation}
  \label{eq:fKfpi_pureQCD}
\frac{f_{K^{\pm}}}{f_{\pi^{\pm}}}=1.1848(59)_{\rm stat}(84)_{\chi-{\rm cont}}(24)_{\rm SU(2)},
\end{equation}
where ``stat'' and ``$\chi-{\rm cont}$'' correspond to the errors in \cref{eq:fKfpi_ph}, i.e. they are uncertainties related to the lattice data, and ``SU(2)'' corresponds to the uncertainty in \cref{eq:delta_SU2_result}.
In \refig{fig:FLAG} we compare our final estimate with results from other collaborations that enter the FLAG average for $N_f=2+1$ \cite{FlavourLatticeAveragingGroupFLAG:2024oxs}.
For the QED correction $\delta_{\rm EM}^{\rm PS}$ to the experimental measurement of \mbox{$\Gamma({\rm PS}\rightarrow l\overline{\nu}_l[\gamma])$}, ${\rm PS}=\pi,K$, we reproduce the same steps as in \cite{Marciano:1993sh,Cirigliano:2007ga}, and we will argue for a conservative choice of the corresponding uncertainty.
The $\delta_{\rm EM}^{\rm PS}$ factor reads
\begin{align}
  \label{eq:delta_EM}
  \delta_{\rm EM}^{\rm PS}&=-\frac{\alpha}{\pi}\left[-F\left(\frac{m_l}{m_{\rm PS}}\right)+\frac{3}{2}\log\left(\frac{m_{\rho}}{m_{\rm PS}}\right)+c_1^{\rm PS}\right. \notag \\
    &\left.+\frac{m_l^2}{m_{\rho}^2}\left(2c_2^{\rm PS}\log\left(\frac{m_{\rho}}{m_l}\right)+c_3^{\rm PS}+c_4^{\rm PS}\frac{m_l}{m_{\rm PS}}\right)-2\tilde{c}_2^{\rm PS}\frac{m_{\rm PS}^2}{m_{\rho}^2}\log\left(\frac{m_{\rho}}{m_{l}}\right)+{\rm h.o.}\right],
\end{align}
where the first term is the universal long distance contribution to one-loop, assuming that the pseudoscalar state PS is point-like, while the terms proportional to $c_i^{\rm PS},\;\tilde{c}_i^{\rm PS}$ parameterise the structure-dependent part.
To the current precision it is enough to consider the $c_1^{\rm PS}$ terms,\footnote{Adding the missing terms, using the results in \cite{Descotes-Genon:2005wrq,Cirigliano:2007ga}, amounts to a shift of a $16\%$ of $\sigma$ in the central value w.r.t. \cref{eq:delta_EM_result}.} whose values are given, using large-$N_c$ methods, by \cite{Ananthanarayan:2004qk,Descotes-Genon:2005wrq,Cirigliano:2011tm}
\begin{align}
  \label{eq:ChPT_EM}
  c_1^{\pi}=-2.5(5), \quad c_1^{K}=-1.9(5).
\end{align}
Since we are only interested in the combination $(1+\delta_{\rm EM}^K-\delta_{\rm EM}^{\pi})$ as evident in \cref{eq:branching}, the relevant quantity to us is
\begin{equation}
  \label{eq:delta_c1}
\Delta c_1\equiv c_1^K-c_1^{\pi}=0.6(6),
\end{equation}
where we took only the central values in \cref{eq:ChPT_EM} and then assumed 100\% uncertainty for $\Delta c_1$, in order to cover the structureless pion and kaon limit in which $\Delta c_1=0$, which we consider to be the most conservative choice.
Taking $l=\mu$ and with the function $F(x)$ explicitly given in \cite{Marciano:1993sh}, we get
\begin{equation}
  \label{eq:delta_EM_result}
\delta_{\rm EM}^{K}-\delta_{\rm EM}^{\pi}=-0.0072(14).
\end{equation}
With the results in \cref{eq:delta_SU2_result} and \cref{eq:delta_EM_result}, as an intermediate comparison with recent lattice computations we quote the joint quantity $\delta R_{K\pi}=\delta_{\rm SU(2)}+\delta_{\rm EM}^{K}-\delta_{\rm EM}^{\pi}$, for which we obtain
\begin{equation}
  \label{eq:RKpi}
\delta R_{K\pi}=-0.0112(40)_{\rm SU(2)}(14)_{\rm EM}.
\end{equation}
The first error comes from that in \cref{eq:delta_SU2_result}, while the second from that in \cref{eq:delta_EM_result}.
The joint quantity $\delta R_{K\pi}$ is of particular importance, as it has been shown that it can be computed from first-principles non-perturbatively on the lattice.
The estimation in \cref{eq:RKpi} is compatible with the most recent lattice determinations of this quantity
\begin{align}
  \label{eq:}
\delta R_{K\pi}&=-0.0126(14) \quad \text{\cite{DiCarlo:2019thl}}, \\
\delta R_{K\pi}&=-0.0086(41) \quad \text{\cite{Boyle:2022lsi}},
\end{align}
but with a more conservative uncertainty.
Despite this fact, as we will see our final result is currently completely dominated by the uncertainty in \cref{eq:fKfpi_ph}.
Using the experimental value \cite{ParticleDataGroup:2024cfk}
\begin{align}
  \label{eq:ratio_exp}
\frac{\Gamma(K\rightarrow l\overline{\nu}_l[\gamma])}{\Gamma(\pi\rightarrow l\overline{\nu}_l[\gamma])}&=1.3367(32)_{\rm exp},
\end{align}
together with the QED correction in \cref{eq:delta_EM_result} and the values for $m_{K^{\pm}},\;m_{\pi^{\pm}},\;m_{\mu}$ reported in \cite{ParticleDataGroup:2024cfk}, following \cref{eq:branching} we arrive at
\begin{align}
\frac{f_{K^{\pm}}|V_{us}|}{f_{\pi^{\pm}}|V_{ud}|}&=0.27604(33)_{\rm exp}(19)_{\rm EM}.
\end{align}
Finally, from our result in \cref{eq:fKfpi_pureQCD}, we obtain
\begin{align}
  \label{eq:Vus/Vud}
  \frac{|V_{us}|}{|V_{ud}|}&=0.2330(11)_{\rm stat}(17)_{\chi-{\rm cont}}(5)_{\rm SU(2)}(2)_{\rm EM}(3)_{\rm exp}.
\end{align}
Employing the value of $|V_{ud}|$ from super-allowed $\beta$-decays $|V_{ud}|=0.97367(32)_{\beta{\rm-dec.}}$ \cite{ParticleDataGroup:2024cfk} and neglecting $|V_{ub}|$, whose value squared is $\mbox{O}(10^{-5})$ \cite{ParticleDataGroup:2024cfk} and thus negligible to the current precision, we obtain
\begin{align}
  \label{eq:}
  |V_{us}|&=0.2269(13)_{\rm stat}(15)_{\chi-{\rm cont}}(4)_{\rm SU(2)}(2)_{\rm EM}(2)_{\beta{\rm-dec.}}(2)_{\rm exp},\\
|V_{ud}|^2+|V_{us}|^2&=0.9995(6)_{\rm stat}(7)_{\chi-{\rm cont}}(2)_{\rm SU(2)}(1)_{\rm EM}(7)_{\beta{\rm-dec.}}(1)_{\rm exp}.
\end{align}
For completeness, we also test employing as input the value $|V_{us}|=0.22328(58)_{|V_{us}|}$, obtained from the experimental determination of $|V_{us}|f_{+}(0)$ \cite{Moulson:2017ive} and the $N_f=2+1+1$ lattice result for $f_{+}(0)$ \cite{FlavourLatticeAveragingGroupFLAG:2024oxs}.
Using this input for $|V_{us}|$ together with our result in \cref{eq:Vus/Vud}, we obtain
\begin{equation}
  \label{eq:}
|V_{ud}|^2+|V_{us}|^2=0.9682(87)_{\rm stat}(134)_{\chi-{\rm cont}}(39)_{\rm SU(2)}(16)_{\rm EM}(50)_{|V_{us}|}(24)_{\rm exp},
\end{equation}
which is at $1.8\sigma$ from the Standard Model unitarity constraint.
If we repeat the same excersise but using now the $N_f=2+1$ lattice result for $f_+(0)$ \cite{FlavourLatticeAveragingGroupFLAG:2024oxs}, then $|V_{us}|=0.22377(75)_{|V_{us}|}$ and we find
\begin{equation}
  \label{eq:}
|V_{ud}|^2+|V_{us}|^2=0.9724(87)_{\rm stat}(180)_{\chi-{\rm cont}}(39)_{\rm SU(2)}(16)_{\rm EM}(65)_{|V_{us}|}(24)_{\rm exp},
\end{equation}
which is at $1.5\sigma$ from unitarity.
We see that our final uncertainty is completely dominated by that of the lattice determination of $f_K/f_{\pi}$ in \cref{eq:fKfpi_ph}, in spite of the conservative choices we took for the uncertainties related to the strong isospin-breaking and QED effects.

\section{Conclusions}
\label{sec:conclusions}

In this work we have presented results for $f_K/f_{\pi}$ in pure isoQCD, in addition to $|V_{us}|/|V_{ud}|$, where strong isospin-breaking and QED effects have been included, which allows to perform a test of the unitarity of the first row of the CKM matrix.
Two important features of our construction are that (i) it is based on a combination of a Wilson unitary action and a mixed action with Wilson twisted mass valence fermions \cite{Bussone:2025wlf,Bussone:2023kag}, and (ii) that it is based on $f_{\pi}$ to set the scale.
This latter point allows to avoid any systematic uncertainty related to the continuum-limit extrapolation of theory scales like $t_0$ or $w_0$.
We quote as our main results
\begin{align}
  \frac{f_K}{f_{\pi}}&=1.1872(59)_{\rm stat}(84)_{\chi-{\rm cont}}[103], \\
  \frac{f_{K^{\pm}}}{f_{\pi^{\pm}}}&=1.1848(59)_{\rm stat}(84)_{\chi-{\rm cont}}(24)_{\rm SU(2)}[105], \\
    \frac{|V_{us}|}{|V_{ud}|}&=0.2330(11)_{\rm stat}(17)_{\chi-{\rm cont}}(5)_{\rm SU(2)}(2)_{\rm EM}(3)_{\rm exp}[21],
\end{align}
the first of which is an unambiguous prediction of $N_f=2+1$ pure isoQCD and can be directly compared to other groups, once the input in \cref{eq:Edinburgh} has been fixed.
Despite the conservative error choices for the strong isospin-breaking effects and QED corrections, we observe that the uncertainty in these results is still completely dominated by the lattice determination of $f_K/f_{\pi}$.
Increased statistics in the ensembles that contribute more to the statistical error in \refig{fig:pie} is expected to help in reducing the error in our results.
With respect to the systematic uncertainty coming from the model variation in the chiral-continuum extrapolations, it is important to note that we have included ensembles as fine as $a\approx0.039$ fm and one (E250, $a\approx0.063$ fm) at the physical pion mass $m_{\pi}\approx130$ MeV.
As pointed out in \res{sec:chiral-continuum}, we find the systematic uncertainty to be dominated by the chiral dependence of $f_K/f_{\pi}$.
Thus, further ensembles at the physical point for finer lattice spacings, in addition to increased statistics in E250, are expected to substantially help in constraining the chiral-continuum extrapolation.
The inclusion of NNLO $\chi$PT terms are also relevant in this respect, for which the inclusion of other chiral trajectories than the ${\rm tr}(M_q)\approx{\rm const.}$ one employed here could be of substantial help.

\section*{Acknowledgments}

The authors want to thank Alberto Ramos for the very useful discussions and critical reading of a previous version of the manuscript.
We also want to thank Gregorio Herdoiza and Carlos Pena for help in the data generation, fruitful discussions and feedback on a previous version of the manuscript.
We are grateful to our colleagues in the CLS initiative for sharing ensembles.
We acknowledge PRACE for awarding us access to MareNostrum at Barcelona Supercomputing Center (BSC), Spain and to HAWK at GCS{\@}HLRS, Germany.
The authors thankfully acknowledge the computer resources at MareNostrum and the technical support provided by Barcelona Supercomputing Center (FI-2020-3-0026).
We thank CESGA for granting access to Finis Terrae II.
We also acknowledge support from the Spanish Research Agency (Agencia Estatal de Investigación) through national project CNS2022-136005, AEI/MCIU through grant PID2023-148162NB-C21 and ASFAE/2022/020.
\begin{appendix}
\section{QCD scheme variation and impact}
\label{app_input}
In this Appendix we test the impact of modifying the definition of isoQCD by taking different values of $m_{\pi,K}^{\rm ph}$ and $f_{\pi}^{\rm ph}$ with respect to those in \cref{eq:Edinburgh}. In particular, we take the values given in \cite{Aoki:2016frl}
\begin{equation}
  \label{eq:DiCarlo_input}
m_{\pi}^{\rm ph}=134.8(3)\;{\rm MeV}, \quad m_{K}^{\rm ph}=494.2(3)\;{\rm MeV}, \quad f_{\pi}^{\rm ph}=130.4(2)\;{\rm MeV}.
\end{equation}
Repeating the analysis in \res{sec:chiral-continuum} with this new input, we see no change in the result in \cref{eq:fKfpi_ph} to the precision obtained.
This is expected since the two schemes are very similar in terms of the numerical values of the input quantities.
For completeness, we also quote the result for the derivatives of $f_K/f_{\pi}$ with respect to the input quantities in \ret{tab:derivatives}.
These derivatives are estimated using the automatic differentiation tools implemented in the \verb|ADerrors| package \cite{Ramos:2018vgu}.
  \begin{longtable}{c c c}
    \toprule
    $\left.\frac{\partial\left(f_K/f_{\pi}\right)}{\partial
    m_{\pi}^{\rm ph}}\frac{m_{\pi}^{\rm ph}}{f_K/f_{\pi}}\right|_{\rm Edinburgh}$ & $\left.\frac{\partial\left(f_K/f_{\pi}\right)}{\partial m_{K}^{\rm ph}}\frac{m_{K}^{\rm ph}}{f_K/f_{\pi}}\right|_{\rm Edinburgh}$ & $\left.\frac{\partial\left(f_K/f_{\pi}\right)}{\partial f_{\pi}^{\rm ph}}\frac{f_{\pi}^{\rm ph}}{f_K/f_{\pi}}\right|_{\rm Edinburgh}$ \\
    \toprule
    $-0.0208$ & $0.4032$ & $-0.3824$ \\
    \bottomrule
    \caption{Derivatives of $f_K/f_{\pi}$ with respect to
    the input quantities $m_{\pi,K}^{\rm ph},\;f_{\pi}^{\rm ph}$ used to define the physical point, which can be
    used to convert our results for $f_K/f_{\pi}$ in \cref{eq:fKfpi_ph} to any
    other scheme defined by different values of $m_{\pi,K}^{\rm ph}$, $f_{\pi}^{\rm ph}$
    in isoQCD.
    The derivatives are evaluated at the isoQCD values given by the Edinburgh Consensus in \cref{eq:Edinburgh}.}
    \label{tab:derivatives}
  \end{longtable}
\end{appendix}

\bibliographystyle{JHEP}
\bibliography{references}

@article{Bussone:2025wlf,
    author = "Bussone, Andrea and Conigli, Alessandro and Frison, Julien and Herdo{\'\i}za, Gregorio and Pena, Carlos and Preti, David and Romero, Jos{\'e} {\'A}ngel and S{\'a}ez, Alejandro and Ugarrio, Javier",
    collaboration = "ALPHA",
    title = "{Hadronic physics from a Wilson fermion mixed-action approach: Setup and scale setting}",
    eprint = "2510.20450",
    archivePrefix = "arXiv",
    primaryClass = "hep-lat",
    reportNumber = "IFT-UAM/CSIC-25-108",
    month = "10",
    year = "2025"
}

@article{Marciano:1993sh,
    author = "Marciano, William J. and Sirlin, A.",
    title = "{Radiative corrections to pi(lepton 2) decays}",
    reportNumber = "NYU-TH-93-09-03",
    doi = "10.1103/PhysRevLett.71.3629",
    journal = "Phys. Rev. Lett.",
    volume = "71",
    pages = "3629--3632",
    year = "1993"
}

@article{Kennedy:1998cu,
    author = "Kennedy, A. D. and Horvath, Ivan and Sint, Stefan",
    editor = "DeGrand, Thomas A. and DeTar, Carleton E. and Sugar, R. and Toussaint, D.",
    title = "{A New exact method for dynamical fermion computations with nonlocal actions}",
    eprint = "hep-lat/9809092",
    archivePrefix = "arXiv",
    doi = "10.1016/S0920-5632(99)85217-7",
    journal = "Nucl. Phys. B Proc. Suppl.",
    volume = "73",
    pages = "834--836",
    year = "1999"
}

@article{Frezzotti:2000nk,
    author = "Frezzotti, Roberto and Grassi, Pietro Antonio and Sint, Stefan and Weisz, Peter",
    collaboration = "Alpha",
    title = "{Lattice QCD with a chirally twisted mass term}",
    eprint = "hep-lat/0101001",
    archivePrefix = "arXiv",
    reportNumber = "CERN-TH-2000-384, MPI-PHT-2000-51, BICOCCA-FT-0027, NYU-TH-00-09-13",
    doi = "10.1088/1126-6708/2001/08/058",
    journal = "JHEP",
    volume = "08",
    pages = "058",
    year = "2001"
}

@article{Clark:2003na,
    author = "Clark, M. A. and Kennedy, A. D.",
    editor = "Aoki, S. and Hashimoto, S. and Ishizuka, N. and Kanaya, K. and Kuramashi, Y.",
    title = "{The RHMC algorithm for two flavors of dynamical staggered fermions}",
    eprint = "hep-lat/0309084",
    archivePrefix = "arXiv",
    doi = "10.1016/S0920-5632(03)02732-4",
    journal = "Nucl. Phys. B Proc. Suppl.",
    volume = "129",
    pages = "850--852",
    year = "2004"
}

@article{Wolff:2003sm,
    author = "Wolff, Ulli",
    collaboration = "ALPHA",
    title = "{Monte Carlo errors with less errors}",
    eprint = "hep-lat/0306017",
    archivePrefix = "arXiv",
    reportNumber = "HU-EP-03-32, SFB-CPP-03-12",
    doi = "10.1016/S0010-4655(03)00467-3",
    journal = "Comput. Phys. Commun.",
    volume = "156",
    pages = "143--153",
    year = "2004",
    note = "[Erratum: Comput.Phys.Commun. 176, 383 (2007)]"
}

@article{Pena:2004gb,
    author = "Pena, Carlos and Sint, Stefan and Vladikas, Anastassios",
    title = "{Twisted mass QCD and lattice approaches to the Delta I = 1/2 rule}",
    eprint = "hep-lat/0405028",
    archivePrefix = "arXiv",
    reportNumber = "DESY-04-079, FTUAM-03-22, IFT-UAM-CSIC-03-31, ROM2F-2004-10",
    doi = "10.1088/1126-6708/2004/09/069",
    journal = "JHEP",
    volume = "09",
    pages = "069",
    year = "2004"
}

@article{Ananthanarayan:2004qk,
    author = "Ananthanarayan, B. and Moussallam, B.",
    title = "{Four-point correlator constraints on electromagnetic chiral parameters and resonance effective Lagrangians}",
    eprint = "hep-ph/0405206",
    archivePrefix = "arXiv",
    reportNumber = "IISC-SHEP-5-04, IPNO-DR-04-05",
    doi = "10.1088/1126-6708/2004/06/047",
    journal = "JHEP",
    volume = "06",
    pages = "047",
    year = "2004"
}

@article{Colangelo:2005cg,
    author = "Colangelo, Gilberto and Fuhrer, Andreas and Haefeli, Christoph",
    editor = "Alexandrou, Constantia and Panagopoulos, Haralambos and Schierholz, Gerrit",
    title = "{The pion and proton mass in finite volume}",
    eprint = "hep-lat/0512002",
    archivePrefix = "arXiv",
    doi = "10.1016/j.nuclphysbps.2006.01.004",
    journal = "Nucl. Phys. B Proc. Suppl.",
    volume = "153",
    pages = "41--48",
    year = "2006"
}

@article{Colangelo:2005gd,
    author = "Colangelo, Gilberto and Durr, Stephan and Haefeli, Christoph",
    title = "{Finite volume effects for meson masses and decay constants}",
    eprint = "hep-lat/0503014",
    archivePrefix = "arXiv",
    doi = "10.1016/j.nuclphysb.2005.05.015",
    journal = "Nucl. Phys. B",
    volume = "721",
    pages = "136--174",
    year = "2005"
}

@article{Descotes-Genon:2005wrq,
    author = "Descotes-Genon, Sebastien and Moussallam, Bachir",
    title = "{Radiative corrections in weak semi-leptonic processes at low energy: A Two-step matching determination}",
    eprint = "hep-ph/0505077",
    archivePrefix = "arXiv",
    reportNumber = "IPNO-DR-05-03, LPT-ORSAY-05-29",
    doi = "10.1140/epjc/s2005-02316-8",
    journal = "Eur. Phys. J. C",
    volume = "42",
    pages = "403--417",
    year = "2005"
}

@article{Clark:2006fx,
    author = "Clark, M. A. and Kennedy, A. D.",
    title = "{Accelerating dynamical fermion computations using the rational hybrid Monte Carlo (RHMC) algorithm with multiple pseudofermion fields}",
    eprint = "hep-lat/0608015",
    archivePrefix = "arXiv",
    doi = "10.1103/PhysRevLett.98.051601",
    journal = "Phys. Rev. Lett.",
    volume = "98",
    pages = "051601",
    year = "2007"
}

@article{Cirigliano:2007ga,
    author = "Cirigliano, Vincenzo and Rosell, Ignasi",
    title = "{pi/K ---\ensuremath{>} e anti-nu(e) branching ratios to O(e**2 p**4) in Chiral Perturbation Theory}",
    eprint = "0707.4464",
    archivePrefix = "arXiv",
    primaryClass = "hep-ph",
    reportNumber = "LAUR-07-4532",
    doi = "10.1088/1126-6708/2007/10/005",
    journal = "JHEP",
    volume = "10",
    pages = "005",
    year = "2007"
}

@article{Follana:2007uv,
    author = "Follana, E. and Davies, C. T. H. and Lepage, G. P. and Shigemitsu, J.",
    collaboration = "HPQCD, UKQCD",
    title = "{High Precision determination of the pi, K, D and D(s) decay constants from lattice QCD}",
    eprint = "0706.1726",
    archivePrefix = "arXiv",
    primaryClass = "hep-lat",
    doi = "10.1103/PhysRevLett.100.062002",
    journal = "Phys. Rev. Lett.",
    volume = "100",
    pages = "062002",
    year = "2008"
}

@article{Luscher:2010iy,
    author = {L\"uscher, Martin},
    title = "{Properties and uses of the Wilson flow in lattice QCD}",
    eprint = "1006.4518",
    archivePrefix = "arXiv",
    primaryClass = "hep-lat",
    reportNumber = "CERN-PH-TH-2010-143",
    doi = "10.1007/JHEP08(2010)071",
    journal = "JHEP",
    volume = "08",
    pages = "071",
    year = "2010",
    note = "[Erratum: JHEP 03, 092 (2014)]"
}

@article{BMW:2010xmi,
    author = "Durr, S. and Fodor, Z. and Hoelbling, C. and Katz, S. D. and Krieg, S. and Kurth, T. and Lellouch, L. and Lippert, T. and Ramos, A. and Szabo, K. K.",
    collaboration = "BMW",
    title = "{The ratio FK/Fpi in QCD}",
    eprint = "1001.4692",
    archivePrefix = "arXiv",
    primaryClass = "hep-lat",
    doi = "10.1103/PhysRevD.81.054507",
    journal = "Phys. Rev. D",
    volume = "81",
    pages = "054507",
    year = "2010"
}

@article{MILC:2010hzw,
    author = "Bazavov, A. and others",
    editor = "Rossi, Giancarlo",
    collaboration = "MILC",
    title = "{Results for light pseudoscalar mesons}",
    eprint = "1012.0868",
    archivePrefix = "arXiv",
    primaryClass = "hep-lat",
    doi = "10.22323/1.105.0074",
    journal = "PoS",
    volume = "LATTICE2010",
    pages = "074",
    year = "2010"
}

@article{Cirigliano:2011tm,
    author = "Cirigliano, Vincenzo and Neufeld, Helmut",
    title = "{A note on isospin violation in Pl2(gamma) decays}",
    eprint = "1102.0563",
    archivePrefix = "arXiv",
    primaryClass = "hep-ph",
    reportNumber = "LA-UR-11-00400, UWTHPH-2011-4",
    doi = "10.1016/j.physletb.2011.04.038",
    journal = "Phys. Lett. B",
    volume = "700",
    pages = "7--10",
    year = "2011"
}

@article{BMW:2012hcm,
    author = {Bors{\'a}nyi, Szabolcs and D{\"u}rr, Stephan and Fodor, Zolt{\'a}n and Hoelbling, Christian and Katz, S{\'a}ndor D. and Krieg, Stefan and Kurth, Thorsten and Lellouch, Laurent and Lippert, Thomas and McNeile, Craig},
    collaboration = "BMW",
    title = "{High-precision scale setting in lattice QCD}",
    eprint = "1203.4469",
    archivePrefix = "arXiv",
    primaryClass = "hep-lat",
    reportNumber = "ITP-BUDAPEST-657, CPT-P004-2012, WUB-12-02",
    doi = "10.1007/JHEP09(2012)010",
    journal = "JHEP",
    volume = "09",
    pages = "010",
    year = "2012"
}

@article{Luscher:2012av,
    author = "Luscher, Martin and Schaefer, Stefan",
    title = "{Lattice QCD with open boundary conditions and twisted-mass reweighting}",
    eprint = "1206.2809",
    archivePrefix = "arXiv",
    primaryClass = "hep-lat",
    reportNumber = "CERN-PH-TH-2012-161",
    doi = "10.1016/j.cpc.2012.10.003",
    journal = "Comput. Phys. Commun.",
    volume = "184",
    pages = "519--528",
    year = "2013"
}

@article{Bruno:2014jqa,
    author = "Bruno, Mattia and others",
    title = "{Simulation of QCD with N$_{f} =$ 2 $+$ 1 flavors of non-perturbatively improved Wilson fermions}",
    eprint = "1411.3982",
    archivePrefix = "arXiv",
    primaryClass = "hep-lat",
    reportNumber = "DESY-14-216, FTUAM-14-48, HIM-2014-01, HU-EP-14-51, MITP-14-091, SFB-CPP-14-89, IFT-UAM-CSIC-14-117",
    doi = "10.1007/JHEP02(2015)043",
    journal = "JHEP",
    volume = "02",
    pages = "043",
    year = "2015"
}

@article{RBC:2014ntl,
    author = "Blum, T. and others",
    collaboration = "RBC, UKQCD",
    title = "{Domain wall QCD with physical quark masses}",
    eprint = "1411.7017",
    archivePrefix = "arXiv",
    primaryClass = "hep-lat",
    reportNumber = "KEK-TH-1769, RBRC-1095, DAMTP-2014-86",
    doi = "10.1103/PhysRevD.93.074505",
    journal = "Phys. Rev. D",
    volume = "93",
    number = "7",
    pages = "074505",
    year = "2016"
}

@article{Durr:2016ulb,
    author = {D\"urr, Stephan and others},
    title = "{Leptonic decay-constant ratio $f_K/f_\pi$ from lattice QCD using 2+1 clover-improved fermion flavors with 2-HEX smearing}",
    eprint = "1601.05998",
    archivePrefix = "arXiv",
    primaryClass = "hep-lat",
    doi = "10.1103/PhysRevD.95.054513",
    journal = "Phys. Rev. D",
    volume = "95",
    number = "5",
    pages = "054513",
    year = "2017"
}

@article{Bruno:2016plf,
    author = "Bruno, Mattia and Korzec, Tomasz and Schaefer, Stefan",
    title = "{Setting the scale for the CLS $2 + 1$ flavor ensembles}",
    eprint = "1608.08900",
    archivePrefix = "arXiv",
    primaryClass = "hep-lat",
    reportNumber = "DESY-16-162, WUB-16-05",
    doi = "10.1103/PhysRevD.95.074504",
    journal = "Phys. Rev. D",
    volume = "95",
    number = "7",
    pages = "074504",
    year = "2017"
}

@article{QCDSF-UKQCD:2016rau,
    author = {Bornyakov, V. G. and Horsley, R. and Nakamura, Y. and Perlt, H. and Pleiter, D. and Rakow, P. E. L. and Schierholz, G. and Schiller, A. and St\"uben, H. and Zanotti, J. M.},
    collaboration = "QCDSF\textendash{}UKQCD",
    title = "{Flavour breaking effects in the pseudoscalar meson decay constants}",
    eprint = "1612.04798",
    archivePrefix = "arXiv",
    primaryClass = "hep-lat",
    reportNumber = "ADP-16-46-T1002, DESY-16-241, EDINBURGH-2016-19, LIVERPOOL-LTH-1116",
    doi = "10.1016/j.physletb.2017.02.018",
    journal = "Phys. Lett. B",
    volume = "767",
    pages = "366--373",
    year = "2017"
}

@article{Mohler:2017wnb,
    author = "Mohler, Daniel and Schaefer, Stefan and Simeth, Jakob",
    editor = "Della Morte, M. and Fritzsch, P. and G\'amiz S\'anchez, E. and Pena Ruano, C.",
    title = "{CLS 2+1 flavor simulations at physical light- and strange-quark masses}",
    eprint = "1712.04884",
    archivePrefix = "arXiv",
    primaryClass = "hep-lat",
    reportNumber = "DESY-17-166, HIM-2017-08, MITP-17-093",
    doi = "10.1051/epjconf/201817502010",
    journal = "EPJ Web Conf.",
    volume = "175",
    pages = "02010",
    year = "2018"
}

@article{Ramos:2018vgu,
    author = "Ramos, Alberto",
    title = "{Automatic differentiation for error analysis of Monte Carlo data}",
    eprint = "1809.01289",
    archivePrefix = "arXiv",
    primaryClass = "hep-lat",
    doi = "10.1016/j.cpc.2018.12.020",
    journal = "Comput. Phys. Commun.",
    volume = "238",
    pages = "19--35",
    year = "2019"
}

@article{DiCarlo:2019thl,
    author = "Di Carlo, M. and Giusti, D. and Lubicz, V. and Martinelli, G. and Sachrajda, C. T. and Sanfilippo, F. and Simula, S. and Tantalo, N.",
    title = "{Light-meson leptonic decay rates in lattice QCD+QED}",
    eprint = "1904.08731",
    archivePrefix = "arXiv",
    primaryClass = "hep-lat",
    doi = "10.1103/PhysRevD.100.034514",
    journal = "Phys. Rev. D",
    volume = "100",
    number = "3",
    pages = "034514",
    year = "2019"
}

@misc{Luscher:2019openQCD,
  author       = {M. Lüscher},
  title        = {Charm and strange quark in openQCD simulations},
  year         = {2019},
  howpublished = {\url{http://luscher.web.cern.ch/luscher/openQCD/}},
}

@article{Boyle:2022lsi,
    author = "Boyle, Peter and others",
    title = "{Isospin-breaking corrections to light-meson leptonic decays from lattice simulations at physical quark masses}",
    eprint = "2211.12865",
    archivePrefix = "arXiv",
    primaryClass = "hep-lat",
    reportNumber = "CERN-TH-2022-193, LU-TP 22-59",
    doi = "10.1007/JHEP02(2023)242",
    journal = "JHEP",
    volume = "02",
    pages = "242",
    year = "2023"
}

@article{Husung:2022kvi,
    author = "Husung, Nikolai",
    title = "{Logarithmic corrections to O(a) and O($a^2$) effects in lattice QCD with Wilson or Ginsparg\textendash{}Wilson quarks}",
    eprint = "2206.03536",
    archivePrefix = "arXiv",
    primaryClass = "hep-lat",
    doi = "10.1140/epjc/s10052-023-11258-8",
    journal = "Eur. Phys. J. C",
    volume = "83",
    number = "2",
    pages = "142",
    year = "2023",
    note = "[Erratum: Eur.Phys.J.C 83, 144 (2023)]"
}

@article{Bruno:2022mfy,
    author = "Bruno, Mattia and Sommer, Rainer",
    title = "{On fits to correlated and auto-correlated data}",
    eprint = "2209.14188",
    archivePrefix = "arXiv",
    primaryClass = "hep-lat",
    doi = "10.1016/j.cpc.2022.108643",
    journal = "Comput. Phys. Commun.",
    volume = "285",
    pages = "108643",
    year = "2023"
}

@article{Neil:2023pgt,
    author = "Neil, Ethan T. and Sitison, Jacob W.",
    title = "{Model averaging approaches to data subset selection}",
    eprint = "2305.19417",
    archivePrefix = "arXiv",
    primaryClass = "stat.ME",
    doi = "10.1103/PhysRevE.108.045308",
    journal = "Phys. Rev. E",
    volume = "108",
    number = "4",
    pages = "045308",
    year = "2023"
}

@article{Bussone:2023kag,
    author = "Bussone, Andrea and Conigli, Alessandro and Frison, Julien and Herdo\'\i{}za, Gregorio and Pena, Carlos and Preti, David and S\'aez, Alejandro and Ugarrio, Javier",
    collaboration = "Alpha",
    title = "{Hadronic physics from a Wilson fermion mixed-action approach: charm quark mass and $D_{(s)}$ meson decay constants}",
    eprint = "2309.14154",
    archivePrefix = "arXiv",
    primaryClass = "hep-lat",
    reportNumber = "IFT-UAM/CSIC-23-114",
    doi = "10.1140/epjc/s10052-024-12816-4",
    journal = "Eur. Phys. J. C",
    volume = "84",
    number = "5",
    pages = "506",
    year = "2024"
}

@article{Kuberski:2023zky,
    author = "Kuberski, Simon",
    title = "{Low-mode deflation for twisted-mass and RHMC reweighting in lattice QCD}",
    eprint = "2306.02385",
    archivePrefix = "arXiv",
    primaryClass = "hep-lat",
    reportNumber = "MITP-23-021",
    doi = "10.1016/j.cpc.2024.109173",
    journal = "Comput. Phys. Commun.",
    volume = "300",
    pages = "109173",
    year = "2024"
}

@article{CLQCD:2023sdb,
    author = "Hu, Zhi-Cheng and others",
    collaboration = "CLQCD",
    title = "{Quark masses and low-energy constants in the continuum from the tadpole-improved clover ensembles}",
    eprint = "2310.00814",
    archivePrefix = "arXiv",
    primaryClass = "hep-lat",
    doi = "10.1103/PhysRevD.109.054507",
    journal = "Phys. Rev. D",
    volume = "109",
    number = "5",
    pages = "054507",
    year = "2024"
}

@article{FlavourLatticeAveragingGroupFLAG:2024oxs,
    author = "Aoki, Y. and others",
    collaboration = "Flavour Lattice Averaging Group (FLAG)",
    title = "{FLAG Review 2024}",
    eprint = "2411.04268",
    archivePrefix = "arXiv",
    primaryClass = "hep-lat",
    reportNumber = "CERN-TH-2024-192, FERMILAB-PUB-24-0785-T",
    month = "11",
    year = "2024"
}

@article{ParticleDataGroup:2024cfk,
    author = "Navas, S. and others",
    collaboration = "Particle Data Group",
    title = "{Review of particle physics}",
    doi = "10.1103/PhysRevD.110.030001",
    journal = "Phys. Rev. D",
    volume = "110",
    number = "3",
    pages = "030001",
    year = "2024"
}

@article{Aoki:2016frl,
    author = "Aoki, S. and others",
    title = "{Review of lattice results concerning low-energy particle physics}",
    eprint = "1607.00299",
    archivePrefix = "arXiv",
    primaryClass = "hep-lat",
    reportNumber = "CP3-Origins-2016-023, DESY-16-111, DIAS-2016-23, Edinburgh-2016-11, FTUAM-16-23, HIM-2016-02, IFT-UAM-CSIC-16-057, LPT-Orsay-16-47, MITP-16-059, RM3-TH-16-7, ROM2F-2016-05, YITP-16-77",
    doi = "10.1140/epjc/s10052-016-4509-7",
    journal = "Eur. Phys. J. C",
    volume = "77",
    number = "2",
    pages = "112",
    year = "2017"
}

@article{Moulson:2017ive,
    author = "Moulson, Matthew",
    title = "{Experimental determination of $V_{us}$ from kaon decays}",
    eprint = "1704.04104",
    archivePrefix = "arXiv",
    primaryClass = "hep-ex",
    doi = "10.22323/1.291.0033",
    journal = "PoS",
    volume = "CKM2016",
    pages = "033",
    year = "2017"
}

@article{Frison:2023lwb,
    author = "Frison, Julien",
    title = "{Towards fully bayesian analyses in Lattice QCD}",
    eprint = "2302.06550",
    archivePrefix = "arXiv",
    primaryClass = "hep-lat",
    month = "2",
    year = "2023"
}

\end{document}